\documentclass{aa}
\usepackage{amssymb,amsmath}
\usepackage{graphicx,epstopdf}
\usepackage{natbib}
\usepackage{bm}
\bibpunct{(}{)}{;}{a}{}{,} 
\usepackage{color}
\def\lesssim{\;\raise0.3ex\hbox{$<$\kern-0.75em\raise-1.1ex\hbox{$\sim$}}\;}
\def\gtrsim{\;\raise0.3ex\hbox{$>$\kern-0.75em\raise-1.1ex\hbox{$\sim$}}\;}
\def\mdens{{\rm g~cm^{-3}}}

\def\msun{{\rm M}_\odot}
\def\macc{{\rm M}_\odot~{\rm yr}^{-1}}
\def\acell{A_{\rm cell}}
\def\dz{\Delta N}
\begin{document}
\title{Crustal heating in accreting neutron stars from the nuclear energy-density functional theory. 
I. Proton shell effects and neutron-matter constraint}
\author{A.~F. Fantina\inst{1,2}, J.~L. Zdunik\inst{3}, 
N. Chamel\inst{2}, J.~M.~Pearson\inst{4}, P. Haensel\inst{3}, 
and S. Goriely\inst{2}}

\institute{Grand Acc\'el\'erateur National d'Ions Lourds (GANIL), CEA/DRF -
 CNRS/IN2P3, Boulevard Henri Becquerel, 14076 Caen, France\\
 \email{anthea.fantina@ganil.fr}
\and Institut d'Astronomie et d'Astrophysique, CP-226, 
Universit\'e Libre de Bruxelles, 1050 Brussels, Belgium \\
\and 
N. Copernicus Astronomical Center, Polish Academy of Sciences, 
Bartycka 18, PL-00-716 Warszawa, Poland\\ 
\and 
D\'ept. de Physique, Universit\'e de Montr\'eal, Montr\'eal
(Qu\'ebec), H3C 3J7 Canada }
         
\date{Received xxx Accepted xxx}

\abstract{Observations of soft X-ray transients in quiescence suggest the 
existence of heat sources in the crust of accreting neutron stars. Heat is thought 
to be released by electroweak and nuclear processes triggered by the burying of 
ashes of X-ray bursts.}
{The heating is studied using a fully 
quantum approach taking consistently into account nuclear shell effects. 
} 
{We have followed the evolution of ashes made of $^{56}$Fe employing 
the nuclear energy-density functional theory. Both the outer and inner crusts are 
described using the same functional, thus ensuring a unified and thermodynamically 
consistent treatment. To assess the role of the neutron-matter constraint, we have 
employed the set of accurately calibrated Brussels-Montreal functionals BSk19, BSk20, 
and BSk21 and for comparison the SLy4 functional. 
}
{Due to nuclear shell effects, the fully accreted crust is found to be 
much less stratified than in previous studies. In particular, large regions of the 
inner crust contain clusters with the magic number $Z=14$. The heat 
deposited in the outer crust is tightly constrained by experimental atomic mass 
data. The shallow heating we obtain does not exceed $0.2$~MeV and is therefore not enough 
to explain the cooling of some soft X-ray transients. The total heat released in the 
crust is very sensitive to details of the nuclear structure and is predicted to lie in 
the range from $1.5$~MeV to $1.7$~MeV. 
}
{The evolution of an accreted matter element and therefore the location of heat sources 
are governed to a large extent by the existence of nuclear shell closures. Ignoring these 
effects in the inner crust, the total heat falls to $\sim 0.6$~MeV. The neutron-matter 
constraint is also found to play a key role. The large amount of heat obtained by Steiner 
et al. (2012) could thus be traced back to unrealistic neutron-matter 
equations of state. 
} 

\keywords{dense matter -- equation of state -- stars: neutron -- accretion -- nuclear reactions}

\titlerunning{Crustal heating in accreting neutron stars}
\authorrunning{A. F. Fantina et al.}

\maketitle
\section{Introduction}
\label{sect:introd}
The crust of a neutron star (NS) contributes only about one percent to the stellar mass. 
However, it is essential for many astrophysical phenomena associated with neutron 
stars, such as pulsar glitches, X-ray bursts, and repetitive gamma-ray flares in 
magnetars. From the dense-matter theory point of view,  the description of the crust 
of a NS with its density below the normal nuclear density, seems much less 
challenging than the core. Still, the properties of the crust beyond the onset of 
neutron drip can only be studied theoretically, because such an  environment cannot 
be reproduced in terrestrial laboratories  (for a review of the physics of NS 
crusts, see, e.g., \cite{lrr}). 

A standard assumption concerning the crust of a non-accreting (isolated) neutron 
star is that it is made of ``cold-catalysed matter'', i.e. electrically charge 
neutral matter at zero temperature in its absolute ground state \citep{Harrison1958,
Harrison1965}. The composition of any crustal layer at pressure $P$ is thus obtained 
from the minimum of the Gibbs free energy per nucleon $g$. However, the interior of 
a NS may not be in full thermodynamic equilibrium. After the formation of 
the star in the aftermath of gravitational core-collapse supernova explosions, strong 
and electroweak nuclear reactions could be quenched due to fast cooling or crystallization. 
Moreover, the composition of the crust could be later altered by the accretion of 
matter from a stellar companion in a close binary system. The crust of a NS 
can thus be a reservoir of nuclear energy, which could be released under favourable 
conditions \citep{Bisno1973,Bisno1974,Bisno1976,Bisno1979,hz1990a,Blaes1990}. 

The energy release in non-equilibrium layers of NS crusts was orginally 
proposed as the potential source of heat powering the thermal X-ray radiation of 
isolated NSs \citep{Bisno1973,Bisno1974,Bisno1976}. Later, the  instabilities in 
non-equilibrium crustal shells were considered as being at the origin of Galactic gamma-ray 
bursts \citep{Bisno1979,Blaes1990}. The models employed in these pioneering studies for 
describing the inner crust, where most of the heat release takes place, were very simplified, 
and therefore the predicted distributions and strengths of the nuclear energy sources were 
very approximate. Systematic calculations of the steady energy release in an accreting 
NS crust were carried out by \cite{hz1990a} (hereafter referred to as HZ), based 
on the compressible liquid-drop model (CLDM) of \cite{mb1977} (hereafter referred to as MB). 
The results of \cite{hz1990a} were applied to the modeling of the thermal structure of accreting 
NSs by \cite{Miralda1990}. 

The interest in the nuclear processes taking place in the crust of accreting NSs grew  in the late 1990s when it was shown that the associated heating could be relevant for 
interpreting the measured surface temperatures of soft X-ray transients (SXTs) in quiescence  \citep{Bildsten1997,Brown1998}. Because the main heat sources predicted by HZ were concentrated in the crustal layers $\sim 300 - 400~$m below the stellar surface, the heat deposition mechanism was thus called "deep crustal heating". The heat is released during accretion
stages (days - weeks). Accretion ({\it bursting}, or {\it active}) stages are characterized by 
X-ray bursts (typically a minute long) separated by about an hour needed to accumulate a new
envelope of nuclear fuel for the next burst. The active periods are separated by much longer
periods of {\it quiescence} with no bursts, and very little accretion or no accretion at all.
For SXTs in quiescence, the thermal radiation from the NS surface can be observed 
and the effective surface temperature can be inferred. 
The crustal heating during active stages is driven by the compression of the deep layers of NS
crust, implying electron captures inducing neutronization and neutron emission, and,  
above $\sim 10^{12}~{\rm g~ cm^{-3}}$, also possibly the pycnonuclear fusion of nuclear 
clusters\footnote{The nuclear clusters constituting the inner crust of a NS differ
radically from ordinary nuclei in that their properties are intimately related to the presence of 
a highly degenerate neutron liquid.}. After $\sim 10^{5}~$yr, a quasiequilibrium quiescent
state is reached with an observable (isotropic) surface photon luminosity $L_X^{\rm (q)}\sim
10^{31} - 10^{33}\;{\rm erg~ s^{-1}}$.  An additional heating $Q_{\rm tot}$ of the deep 
layers of the crust during active periods, by some $1.5~$MeV per accreted nucleon (nicely
consistent with HZ), together with data referring to the overall  time-averaged accretion 
rate for a given SXT, reproduce most of the measured values of  $L_X^{\rm (q)}$ for two 
dozen of SXTs, with some remarkable exceptions (\cite{Beznogov2015} and references therein). 

The HZ model was improved and extended in subsequent works on crustal heating in the framework of the CLDM. \cite{HZ2003} showed, using the MB model, that $Q_{\rm tot}$ is only  weakly dependent on the composition of the X-ray burst ashes. \cite{Gupta2007} considered a network of nuclear
reactions in a hot multicomponent plasma of nuclei taking into account excited states, 
but their study was restricted to the outer crust only. Nuclear masses were obtained  
from the finite-range droplet model of \cite{frdm1995}, whereas the pressure of free 
neutrons was calculated using the MB model. The equation of state (EoS) of electrons 
(including electron-ion electrostatic corrections) was taken from \cite{timmes2000}. 
\cite{HZ2008} later showed that the one-component plasma approximation at $T=0$, as 
implemented in the MB model, leads to a similar estimate for the cumulated heat 
$Q(\rho)$ (total crustal heating per one accreted nucleon integrated up to density $\rho$) as that 
obtained by \cite{Gupta2007}, provided neutrino losses are ignored 
thus mimicking in this way the effect of the de-excitation of the final states of the 
daughter nuclei in the electron captures. \cite{Gupta2008} followed the evolution of accreted material down to the shallowest layers of the inner crust at densities $\sim 10^{12}$~g~cm$^{-3}$ including neutron captures and  dissociations. \cite{steiner2012} later calculated  
crustal heating in a multicomponent plasma in quasistatistical equilibrium, using a set of CLDM that
include an empirical nuclear shell correction. The total heat released $Q_{\rm tot}$ 
was found to be significantly higher than that calculated in previous studies. More recently, \cite{Schatz2014} studied the role of $\beta^-$ decay in the crust reactions and showed that the Urca process can occur in some layers. In a very  recent paper \cite{LauBeard2018} applied the same crust model as \cite{Gupta2007,Schatz2014}, but they considered various compositions of initial ashes, and a large network of reactions including electron captures, $\beta^-$ decays, neutron captures and dissociations as well as the first pycnocnuclear reactions. They found a  total heating 
$Q_{\rm tot}\sim 2~{\rm MeV}/{\rm nucleon}$ quite independent of the initial composition of ashes. However, they could not extend their calculations beyond $2\times 10^{12}~\mdens$ because their model
of nuclei \citep{frdm1995} does not take into account the effect of the surrounding neutron liquid in the inner crust.

The aim of this paper is to reexamine crustal heating in accreted NS crusts, and in 
particular to clarify the role of nuclear shell effects on the location of the heat 
sources as well as on the total heat released. To this end, we follow a more microscopic 
approach based on the self-consistent nuclear energy-density functional (EDF) theory 
\citep{bhr03,sto07}. Unlike the work of \cite{Gupta2007,steiner2012,LauBeard2018}, we adopt the 
one-component plasma approximation, which was shown to be accurate enough for the calculation 
of the cumulated heat. The Brussels-Montreal EDFs BSk19, BSk20, and BSk21 that we employ 
here have been fitted to the same wealth of nuclear data and to different neutron-matter EoS 
based on many-body methods using realistic nuclear forces. Moreover, unified EoSs of catalysed 
matter have already been calculated for these EDFs \citep{fantina2013} thus allowing for a 
direct comparison between nonaccreted and accreted NS crusts. However, in order to better 
assess the sensitivity of our predictions with respect to nuclear physics 
uncertainties, we shall also consider the SLy4 EDF \citep{chabanat1998} that underlies the popular SLy unified EoS \citep{douchinhaensel2001}. 

The plan of the paper is as follows. In Sect.\ref{sect:formation} we review the 
astrophysical scenarios of formation of NS crusts, the difficulties, and the 
limitations of the calculation of crustal heating. Our microscopic model of accreted crust 
is presented in  Sect.\ref{sect:micro}. The evolution of an element of crust matter and  
its composition as a function of the matter density are described in 
Sect.\ref{sect:acc.crust.composition}. In Sect.\ref{sect:acc.crust.heating}, the distribution 
of crustal heating sources is studied for various approximations for the energy of the Wigner-Seitz (WS) cell 
is described and the importance of the shell correction to the energy is illustrated.  We also 
illustrate there the importance of using EDFs that are consistent with 
up-to-date nuclear constraints, both of the theoretical microscopic origin, and 
semi-empirical ones. Discussion, including a comparison with previous 
calculations, and interpretation of generic features of our results, and their 
relation to basic nuclear matter parameters, is  presented in 
Sect.\ref{sect:discussion}.

\section{Formation of accreted crust}
\label{sect:formation}
\subsection{Fully accreted crust vs. partially accreted crust}
Let us consider the evolution of the crust starting from the very beginning of 
the mass transfer stage in low-mass X-ray binaries (LMXB). The history of a SXT is 
a succession of two intermittent stages: periods when matter from the accretion disk 
falls onto the NS ($t_{\rm a})$, interceded by quiescence periods $t_{\rm q}\gg 
t_{\rm a}$ with no or very little accretion onto the NS. Active periods are 
characterised by observations of X-ray bursts. 
This is explained as follows.  During an 
active period, matter is being accreted onto the NS surface, at some rate 
$\dot{M}_{\rm a}$, which can be calculated from the X-ray luminosity of the NS 
surface between X-ray bursts. Typically for SXTs, $\dot{M}_{\rm a}\sim 
10^{-10}-10^{-9}\;\macc$. Accreted matter is hydrogen rich and settles on the NS 
surface, forming an envelope with a hydrogen burning shell at $\sim 
10^{5}\;\mdens$. Below the hydrogen burning shell, helium is accumulating and ignites 
after reaching critical conditions for runaway thermonuclear burning. A 
thermonuclear explosion burns all the overlying envelope within a second, 
leading to the brightening of the NS surface corresponding to an X-ray burst of 
isotropic luminosity $\sim 10^{38}~{\rm erg~s^{-1}}$, which is decaying and 
softening its spectrum on a timescale of a minute. The thermonuclear burning of 
the envelope produces a layer of nuclear ashes, composed of the iron peak and 
heavier nuclides. The next X-ray burst is preceded by an accretion stage  lasting 
from hours to days, and this sequence of X-ray bursts separated by nuclear fuel 
cumulation periods  is continued during  $t_{\rm a}$ (typically weeks - months). 
During quiescent periods, the accretion rate is so low or just zero that in spite 
of $t_{\rm q}\gg t_{\rm a}$ an amount of thermonuclear fuel which can be 
cumulated (if any)  is not sufficient to trigger even a single burst.  The overall 
time averaged accretion rate is 
\begin{equation}
\langle \dot{M} \rangle =\frac{t_{\rm a}}{t_{\rm a}+t_{\rm q}}\;\dot{M}_{\rm 
a}\simeq 
\frac{t_{\rm a}}{t_{\rm q}}\;\dot{M}_{\rm a}~.
\label{eq:acc-rates}
\end{equation}
Here, accretion rates refer to the baryon (rest) mass of the matter, and time is 
measured by a distant observer. 
In view of the strict conservation of the baryon number, the accretion of a baryon 
mass $\Delta M_{\rm a}=t_{\rm a}\langle \dot{M}\rangle$ implies the replacement of 
the ``old'' outer layer of baryon mass $\Delta M_{\rm a}$ by accreted and possibly 
processed material. Let us denote the initial baryon mass of the crust by 
$M^{(\rm i)}_{\rm b,cr}$. After time $t$, the top layer of the crust will be 
replaced  by a mass $t\langle \dot{M}\rangle$ of accreted matter. The same amount
of baryon mass of the original crust will be pushed into the liquid core, 
and converted into a homogeneous plasma phase in a timescale 
 $t_{\rm FA}=M^{(\rm i)}_{\rm b,cr}/\langle \dot{M}\rangle$. However, as long as $t<t_{\rm FA}$, 
only the outer parts of the original crust will be replaced by accreted material. 
We shall refer to these two situations as \emph{partially accreted crust} and 
\emph{fully accreted crust} respectively. 
\subsection{Timescales}
For $t\gg t_{\rm q}$ the accreted  crust mass is given by
\begin{equation}
M^{\rm cr}_{\rm acc}=t\langle \dot{M}\rangle~.
\label{eq:cr.acc}
\end{equation}
The replacement of the original outer crust by accreted matter requires the 
accretion of $10^{-4.6}\;\msun$ (see Fig.36 in \cite{lrr}), which takes
\begin{equation}
t_{\rm oc}=10^{5.4}\;{\rm yr}/\langle \dot{M}\rangle_{-10}~,
\label{eq:acc.oc}
\end{equation}
where $\langle \dot{M}\rangle_{-10}= {\langle \dot{M}\rangle/10^{-10}\;\macc}$. 
After $t_{\rm oc}$, the outer crust acquires a quasistatic structure with a 
steady heating during the active stages.  

However, the most powerful crustal heating comes from the inner layer of the crust at 
densities between $10^{11.5}\;\mdens$ and $10^{13}\;\mdens$. To replace  this 
crucial layer, the accretion of  $10^{-3.6}\;\msun$ is 
needed (see Fig.36 in \cite{lrr}). This takes 
\begin{equation}
t_{\rm DCH}=\frac{10^{6.4}\;{\rm yr}}{\langle \dot{M}\rangle_{-10}}~.
\label{eq:acc.ic}
\end{equation}
Basically, $t_{\rm DCH}$ is the time needed  to obtain a nearly full crustal heating.  After 
$t_{\rm DCH}$,  the partially accreted crust heating regime is left, and crustal heating is 
well approximated by a fully accreted crust model. This stems from the fact that 
the integrated heat $Q(\rho)$ saturates at  $10^{13}~\mdens$ 
(e.g., \cite{HZ2008}, see also Sect.\ref{sect:acc.crust.heating}).   

It should be mentioned that the time needed  for the formation of a fully 
accreted crust at the same mean accretion rate is much longer, $t_{\rm 
FA}\approx 40  t_{\rm DCH}= 10^{8}\;{\rm yr}/\langle \dot{M}\rangle_{-10}~$.

\subsection{Astrophysical context}
\label{sect:astro.context}
Let us consider the 24 SXTs listed in Table 2 in \cite{Beznogov2015}, where references for 
the SXTs data are also given.  For 10 of these transients, $\langle 
\dot{M}\rangle$ has been determined, and ranges between $2.5\times 
10^{-12}\;\macc$ and 
$4\times 10^{-10}\;\macc$. For the remaining SXTs only upper bounds to $\langle 
\dot{M}\rangle$ were established, and they range from  $3\times 10^{-12}\;\macc$ 
to $5\times 10^{-9}\;\macc$. 

In view of the expected lifetime of a SXT, of order $10^8 - 10^9\;$yr, we conclude 
that SXT with 
$\langle \dot{M}\rangle \gtrsim  10^{-10}\;\macc$ could reach the stage of 
nearly full crustal heating. This refers to Aql X-1, 4U~1608$-$522, MXB 1659$-$29, NGC 6440 
X-1, RXJ1709$-$2639, and Terzan 5. On the other hand, XRTs with 
$\langle \dot{M}\rangle \lesssim  10^{-12}\;\macc$ could not reach full power of 
the crustal heating in their partially accreted crusts. This is the case of 
 IGR~00291$+$5934, SAXJ11808.4$-$3658,  XTEJ1751$-$305, and XTE J~1814$-$338. 

In the following, we shall calculate the full crustal heating.

\section{Microscopic model of accreted crust}
\label{sect:micro}

The crust of an accreting NS is covered by an envelope, whose composition 
depends on the NS history and can be strongly time dependent (in X-ray 
bursters). The envelope is crucial for the spectrum of photons  emitted by NS, 
and for the transport of heat between the NS core and surface 
\citep{Potekhin2003}. In the present paper the bottom density of the envelope will 
be fixed at $10^8~\mdens$, and by the NS crust we will mean the layers with 
density higher than this value. 

\subsection{Main assumptions} 

Under the conditions generally prevailing in accreting NS crusts, namely 
$T<10^8$~K (see, e.g. \cite{lrr}), the thermal contributions to the thermodynamic potentials can be neglected. Each crustal layer will be assumed to contain structures made of only one type of 
nuclides with proton number $Z$ and mass number $A$. For simplicity, we further suppose that the crust consists of an 
ordered solid, as suggested by the analysis of the SXT data~\citep[see e.g.][]{lrr}. The structure of each layer is therefore 
fully determined by the composition of a single spherical WS cell. As in \cite{HZ2008}, we shall only consider ground-state transitions and we shall neglect neutrino losses. These approximations will be used throughout this 
paper. 

In hydrostatic equilibrium, the pressure $P$ throughout the star must vary continuously; therefore, the suitable thermodynamic potential for determining the equilibrium structure of the crust is 
the Gibbs free energy per nucleon $g(A,Z,P)$.

\subsection{Evolution of an element of accreted matter}

The determination of the composition of the accreted crust and the sources of crustal heating follows the same approach as in \cite{HZ2008}. 

Let us consider the evolution of some matter element, composed of nuclei $(A,Z)$ compressed by accretion. 
Before the onset of pycnonuclear reactions, the mass number $A$ remains unchanged as the pressure increases. 
On the other hand, the proton number $Z$ can change due to the capture of an electron (simultaneous multiple electron captures 
are very unlikely and are therefore not considered): the nucleus $(A,Z)$ transforms into a nucleus $(A,Z-1)$ with 
proton number $Z-1$ and mass number $A$ with the emission of an electron neutrino. This reaction occurs 
as soon as the pressure reaches a threshold value $P_\beta$ such that $g(A,Z,P_\beta)=g(A,Z-1,P_\beta)$. 
Very accurate analytical formulas for $P_\beta$ and the corresponding threshold density $\rho_\beta$ were obtained by 
\cite{chamel2015b,chamel2016}. 
The electron capture will generally be almost immediately followed by further electron captures on the daughter 
nucleus $(A,Z-1)$. This chain of reactions stops at $Z_0$ defined by $g(A,Z_0,P_\beta)<g(A,Z_0-1,P_\beta)$, which 
corresponds to a \emph{local} minimum of the Gibbs free energy per nucleon at pressure $P$. The \emph{global} minimum 
of $g$ (obtained by relaxing the condition of fixed $A$) defines the state of cold catalysed matter in complete thermodynamical 
equilibrium. As discussed by \citet{HZ2008}, the heat effectively deposited in matter per one WS cell for the 
chain of reactions $(A,Z)\rightarrow (A,Z_0)$ is approximately 
given by 
\begin{equation}
\label{eq:heat-released}
Q_\textrm{cell}\approx  \left[g(A,Z,P_\beta)-g(A,Z_0,P_\beta)\right] A \, .  
\end{equation} 

With increasing pressure, nuclei become progressively more neutron-rich due to electron captures. At some point, 
the nucleus $(A,Z)$ may become unstable against the capture of electrons accompanied by the emission of free 
neutrons. The nucleus $(A,Z)$ will thus transform into a nucleus $(A-\Delta N,Z-1)$ with 
proton number $Z-1$ and mass number $A-\Delta N$ with the emission of $\Delta N$ neutrons and an electron 
neutrino. As previously discussed by \cite{chamel2015}, the mean-nucleus approach leads to a discontinuous variation of the neutron density and of the neutron chemical potential due to the sudden appearance of $\Delta N$ free neutrons in each WS cell. Such unphysical jumps can also be seen in the results obtained by \cite{steiner2012} from a multicomponent CLDM (see his Figs. 3 and 4). Considering that only an infinitesimal small fraction of nuclei $(A,Z)$ are initially converted into $(A-\Delta N,Z-1)$, \cite{chamel2015} showed that the onset of neutron emission actually occurs at lower density and pressure than those predicted by the mean-nucleus treatment. We shall follow the same approach here. In this way, the neutron-drip transition is guaranteed to be continuous, as found in full reaction network calculations~\citep{Gupta2008,LauBeard2018}.

The number $\Delta N$ of emitted neutrons may be larger than one depending on the composition 
of the X-ray burst ashes and on the nuclear mass model employed \citep{chamel2015,fantina2016}. The neutron-drip transition 
delimits the boundary between the outer crust and the inner crust, where neutron-proton clusters coexist with a 
neutron liquid and a gas of electrons. Electron captures (with neutron emission and absorption), pycnonuclear reactions, and crustal 
heat in the inner crust are determined as in the outer crust, except that $A$ is now replaced by the total number 
$A_{\rm cell}$ of nucleons in the WS cell. 

As the matter element sinks deeper inside the star and undergoes further electron captures 
thus decreasing $Z$, the Coulomb barrier  between nuclei may become low enough, and 
the energy of the zero-point vibrations around the lattice sites high enough to trigger
 pycnonuclear reactions, where nuclei $(A,Z)$ fuse by penetrating the Coulomb barrier into 
 nuclei $(2A,2Z)$. However, in view of the large uncertainties in the rates of these reactions 
\citep[see e.g.][]{yak06b}, we simply assume that these processes occur whenever $Z$ reaches the minimum value $Z_\textrm{min}=8$. 
We checked this assumption. For this value of $Z$, the timescale for first pycnonuclear  O-O  fusion at  $1.2\times 10^{12}~\mdens$ 
is estimated as $\tau_{\rm pyc}(Z=8)\sim 0.1~{\rm s}\ll \tau_{\rm DCH}$, 
whereas for  Si-Si, predicted to be produced in 
the O-O fusion and subsequent electron captures, $\tau_{\rm pyc}(Z=14)\sim 10^{30}~{\rm s}\gg \tau_{\rm DCH}$. 
However, further compression of  Si layer induces electron 
captures on Si nuclei and neutron emissions,  leading  again to $Z=8$  and O-O pycnonuclear 
fusion (see Tables A.1 - A.3). 
We used formulae for the pycnonuclear fusion rate derived in the classical paper of 
\cite{SVH1969}, as adapted for accreting NS by \cite{Sato1979}. The formula for the 
astrophysical S-factor, was taken from \cite{Sato1979}. We checked that using up-to-date 
theoretically calculated S-factors calculated in \cite{Afanasjev2012} does not essentially 
change our results, because of the dominating dependence of the fusion rate on $Z$.
 We neglected thermal enhancement of the fusion rate, studied in \cite{yak06a}, because
 for our processes the Coulomb barrier is significantly higher (Si-Si  and O-O  
 instead of C-C and O-C in \cite{yak06a})  and the density is   much 
 larger ($10^{12}~\mdens$ instead 
 of $10^{10}~\mdens$  in \cite{yak06a}).

Our treatment of accreted crust relies on the EDF theory. For the sake of comparison, we shall also consider an improved version of the MB model employed previously by HZ. 

\subsection{Nuclear energy-density functional theory}
\label{subsect:EDF}
Our models of accreting NS crusts are based on the nuclear EDF theory~\citep{bhr03,sto07}. 
This theory, which has proved to be very successful for 
describing the properties of finite nuclei such as those encountered in the outer crust 
of a NS, also allows for a consistent treatment of the inner crust, where neutron-proton 
clusters coexist with a neutron liquid. Moreover, the EDF theory can be applied to homogeneous 
nuclear matter thus providing a unified description of all regions of a NS. The structure, 
the composition and the EoS of nonaccreted NSs was determined in this way 
\citep[see][]{pearson2011,pearson2012} under the assumption of cold catalysed matter. Analytical representations 
of these EoSs were obtained by \citet{potekhin2013}. Using the same 
Brussels-Montreal EDFs labeled BSk19, BSk20 and BSk21 respectively, we shall now study the 
properties of accreting NS crusts.

The Brussels-Montreal EDFs that we consider here are based on generalised Skyrme effective 
nucleon-nucleon interactions \citep{chamel2009, goriely2010}, supplemented with a microscopic 
contact pairing interaction \citep{chamel2010}. These EDFs were fitted to the 2149 measured masses 
of nuclei with neutron and proton numbers, $N$ and $Z \geq 8$ respectively, given in the 2003 
Atomic Mass Evaluation (AME) \citep{audi2003}, with 
a root-mean square (rms) deviation as low as 0.58 MeV for the three functionals, and an optimal 
fit to charge radii. The masses of bound nuclei were obtained by adding to the Hartree-Fock-Bogoliubov (HFB) energy a phenomenological Wigner term and correction term for the collective energy. The precise 
fit to the mass data 
thus makes the functionals well-suited for the description of the nuclei found in the outer crust 
of a NS. The symmetry coefficient was set at $J$ = 30 MeV for the three functionals 
\citep[see e.g.][for a more thorough discussion of experimental and theoretical estimates 
of $J$]{goriely2010, tsang2012, tews2017}. This value of the symmetry energy was later 
found to yield the best fit to nuclear masses for the EDFs in \citet{goriely2013}, and is also supported by NS observations~\citep{pearson2014}. Furthermore, \textit{(i)} the incompressibility 
$K_v$ of symmetric nuclear matter at saturation was required to fall in the experimental range 
$240\pm10$~MeV~\citep{col04}, \textit{(ii)} the ratio of the isoscalar effective mass to bare 
nucleon mass in symmetric nuclear matter at saturation was set to the realistic value of 0.8 
\citep[see\ the\ discussion\ by][]{goriely2003}, \textit{(iii)} the isovector effective mass 
was found to be smaller than the isoscalar effective mass, in agreement with both experiments 
and many-body calculations \citep[see ][for a detailed discussion]{goriely2010}, \textit{(iv)} 
a qualitatively realistic distribution of the potential energy among the four spin-isospin channels 
in nuclear matter was obtained, and \textit{(v)} spurious spin and spin-isospin instabilities 
in nuclear matter that generally plague earlier Skyrme functionals have been eliminated for all 
densities prevailing in NSs by extending the Skyrme functional \citep{chamel2009, chamelgoriely2010}. 
The Brussels-Montreal EDFs BSk19, BSk20 and BSk21 were also constrained to reproduce various 
properties of homogeneous nuclear matter as obtained from many-body calculations using realistic 
two- and three- nucleon interactions. In particular, these EDFs were fitted to three different 
neutron-matter EoSs, reflecting the current lack of knowledge of the high-density behaviour of 
dense matter. BSk19 was adjusted to the ``soft'' EoS of neutron matter of \citet{fp1981} obtained 
from the realistic Urbana v14 nucleon-nucleon force with the three-body force TNI, BSk20 was fitted 
to the EoS of \citet{apr1998} labelled ``A18 + $\delta v$ + UIX'', whereas BSk21 was constrained to 
reproduce the ``stiff'' EoS labelled ``V18'' from \citet{lischulze2008}. These EoSs are compatible 
with the constraints inferred from more recent calculations based on auxiliary field diffusion Monte 
Carlo method and chiral effective field theory \citep{fantina2014}. With these features, the EDFs BSk19, 
BSk20, and BSk21 are well-suited for describing the crust of accreting  NSs. 

For comparison, we shall also consider the SLy4 EDF~\citep{chabanat1998}, which was employed by \citet{douchinhaensel2001} 
 to compute the EoS of the inner crust and core of nonaccreted NS within a CLDM. The SLy4 EDF was also adopted by \cite{steiner2012} for the calculations of accreted NS crusts. Contrary to the Brussels-Montreal EDFs, the SLy4 EDF was not fitted to a particular realistic neutron-matter EoS, but was only constrained to yield a ``reasonable reproduction'' of the EoS of \citet{wir88} up to a density of  1.6 fm$^{-3}$. Moreover, the SLy4 EDF was fitted to only five atomic masses ($^{40,48}$Ca, $^{56}$Ni, $^{132}$Sn, and $^{208}$Pb) from a modified version of AME 1995 \citep{audi1997}. By including $N=Z$ nuclei 
 in the fit without adding a Wigner term, the SLy4 functional overestimates the symmetry coefficient. 
 The rms deviation from the 2003 AME data \citep{audi2003} (considering only even-even nuclei) is about 5.1 MeV~\citep{dsn04}, about an order of magnitude larger than for the Brussels-Montreal models. Although the Rs and Gs EDFs~\citep{fried86} were considered by \cite{steiner2012}, we will not include them here since these EDFs yield unrealistic neutron-matter EoS, as we shall discuss in Section~\ref{subsect:EOS-PNM}. 
 
The EDF theory is implemented using different methods in the outer and inner crusts, as explained in the following sections. 

\subsection{Outer crust}
\label{sect:outer.crust.model}

The properties of the outer crust are calculated in the framework of the standard model of \cite{bps1971}.
This model assumes that the outer crust is made of fully ionised atoms arranged in a body-centred cubic 
lattice neutralised by a degenerate electron gas. 
We follow the same approach as in \cite{pearson2011}, except that we now include electron charge polarization effects using Eqs.(6) and (8) from  \cite{chamel2016b}.

The only microscopic inputs of this model are the \emph{nuclear} masses $M^\prime(A,Z)$, which were obtained from the available 
experimental \emph{atomic} masses $M(A,Z)$ 
after subtracting out the binding energy of 
atomic electrons using Eq.~(A4) of \cite{lpt03}.  
For the masses that have not yet been measured, we used the Brussels-Montreal HFB atomic 
mass tables from the BRUSLIB database \citep{bruslib}. 
Because the SLy4 mass table of \citet{dsn04} contain only even-even nuclei, we have employed the full SLy4 mass table recalculated for the purpose of this work within the HFB method, using the same framework as described in the original paper.

The improvements of this model compared to the previous works of \citet{hz1990a, hz1990b,HZ2003,HZ2008} and \citet{steiner2012} 
are twofold: ({\it i}) the use of more recent atomic mass measurements, and ({\it ii}) the use of much more microscopic and more 
precise nuclear models to predict experimentally unknown nuclear masses. In particular, experimental nuclear 
masses are taken from the 2016 AME \citep{audi2016}, 
whereas \citet{hz1990a, hz1990b,HZ2003,HZ2008} used data from 
the 1971 AME \citep{wapstra71}. Nuclear shell effects, which were not included in the MB model employed by 
\citet{hz1990a, hz1990b,HZ2003,HZ2008}, are now calculated microscopically. Moreover, the CLDM of \citet{hz1990a, hz1990b,HZ2003,HZ2008} 
assumes that nucleons are uniformly distributed within the nucleus with a sharp surface. For the HFB mass tables 
employed in the present work, the nucleon distributions were calculated fully self-consistently and quantum-mechanically 
using the EDF theory. Not surprisingly, the MB model reproduced rather poorly the nuclear mass data available at that 
time, with a rms deviation of 2.6 MeV for 1126 masses of heavy nuclei with $A\geq 40$. \citet{steiner2012} developed a series of CLDMs fitted to the 2003 AME data \citep{audi2003} with a rms error of 1.2 MeV. The relatively low rms error compared to typical CLDM was achieved by introducing phenomenological 
shell corrections. The Brussels-Montreal HFB nuclear 
mass models employed in this work were fitted to the 2149 measured masses of nuclei with $N$ and $Z \geq 8$ given in the 2003 AME \citep{audi2003} 
with a rms deviation lying below $0.6$ MeV. These models also yield an equally good fit to the 2408 experimental masses of nuclei with $Z,N\geq 8$ from the 2016 AME \citep{audi2016}. 
For comparison, the HFB mass table based on SLy4 and the same pairing force as BSk19-21 that we calculated yields a rather poor fit to the same data, with a rms deviation of 3.961 MeV, and a relatively bad description of the closed shell nuclei with respect to open-shell ones due to the low value of 0.69 for the isoscalar effective mass.

\subsection{Neutron-drip transition}
\label{subsec:n-drip}

The onset of neutron emission by nuclei is determined as discussed by \cite{chamel2015}, thus ensuring the thermodynamic consistency across the boundary between the outer and inner crusts. In particular, the neutron chemical potential varies continuously across the transition contrary to the results shown in Figs.~3 and 4 in \cite{steiner2012}. Very accurate analytical formulas for the neutron-drip pressure and density can be found in \cite{chamel2016}. 

As shown by \cite{chamel2015}, the atomic number $Z$ of the dripping nucleus is given by the highest number of protons lying below that of the ashes for which the $\Delta N$-neutron separation energy is negative, or equivalently
\begin{equation}
M^\prime(A-\Delta N,Z-1) < M^\prime(A,Z-1) + \Delta N m_n  \, .                       
\end{equation}
The daughter nucleus may undergo further electron captures and neutron emisions.

\subsection{Inner crust}
\label{sect:inner.crust.model}

Because neutrons are highly degenerate, they contribute to the pressure and can substantially affect the mass of the clusters. 
However, fully self-consistent calculations of the inner crust within the EDF theory are computationally extremely costly. For this 
reason, we have implemented a computationally high-speed approximation based on the 4th order Extended Thomas-Fermi (ETF) development with 
proton shell corrections added perturbatively via the Strutinsky integral theorem~\citep{onsi2008,pearson2012}. Neutron shell corrections, which 
were shown to be much smaller than proton shell corrections~\citep{oya94}, are neglected. The validity of this Extended Thomas-Fermi
with Strutinsky Integral (ETFSI) method has been discussed by \citet{pearson2012}. 
We have employed the same EDFs as those underlying the HFB nuclear mass models used 
for the outer crust, thus ensuing a consistent description of the outer and inner parts of the crust.  

Although neutrons and protons may arrange themselves 
into so called nuclear ``pastas'', these configurations, if they exist, are only expected to be found near the crust-core 
interface~\citep[see e.g. Section 3.3 in][for a brief review]{lrr}. Therefore, nuclear clusters are assumed to be spherical 
and the Coulomb energy is calculated using the WS approximation. In order to further reduce the computation 
time, the nucleon density distributions in the WS cell are parameterised as follows ($q = n,p$ for neutrons, protons respectively): 
\begin{equation}
n_q (r) = n_{B q} + n_{\Lambda q} f_q (r)\, , 
\end{equation}
where $r$ is the radial distance from the centre of the spherical cell, $n_{B q}$ is the background density while the second 
term accounts for the presence of inhomogeneities, with the dimensionless function $f_q(r)$ given by 
\begin{equation}
f_q(r) = \frac{1}{1 + \exp \left[ \Big(\frac{C_q - R}
{r - R}\Big)^2 - 1\right] \exp \Big(\frac{r-C_q}{a_q}\Big) }\, ,
\end{equation}
$R$ being the radius of the WS cell. The expression of $f_q(r)$ ensures that density derivatives vanish at the surface of 
the cell, thus providing a smooth matching of the nucleon distributions between adjacent cells. 

In the absence of pycnonuclear reactions, $A_{\rm cell}$ remains unchanged with increasing pressure whereas $Z$ can decrease if 
electrons are captured. The configuration of a matter element at any pressure $P$ can thus be determined by minimising the baryon 
chemical potential $g$ with respect to all the parameters of the WS cell keeping $A_{\rm cell}$ fixed. However, this procedure 
would be computationally very costly. Instead, we first minimised the internal energy $e$ per nucleon at constant average baryon 
number density $n$, and for given values of the proton number $Z$ and nucleon number $A_\textrm{cell}$ (details can be found in \citealt{pearson2012}). In a second stage, we calculated the corresponding pressure $P$ and the Gibbs free energy per nucleon 
$g$. We repeated this calculation for different values of $n$ and $Z$. In this way, we evaluated $g$ as a function of $Z$ and $P$ 
for the given value of $A_\textrm{cell}$.

Neutrons resulting from electron captures by nuclear clusters may be either bound or unbound. Contrary to the CLDM used by 
\cite{hz1990a, hz1990b,HZ2003,HZ2008} and \citet{steiner2012}, we do not need to consider these two cases separately since both 
neutrons in clusters and free neutrons are treated consistently in the EDF theory. The number $\Delta N$ of emitted neutrons 
can thus take any real positive value. For the sake of comparison, it is convenient to introduce the neutron-cluster number 
$N$ defined as 
\begin{equation}
N = 4\pi\,n_{\Lambda n}\int_0^R r^2f_n(r)dr \,   .
\end{equation}
Another improvement 
compared to the work of \cite{hz1990a, hz1990b,HZ2003,HZ2008} is the inclusion of proton shell effects. As mentioned earlier, these effects were 
considered by \citet{steiner2012} but using an empirical parametrization, whereas shell effects are consistently and quantum mechanically 
calculated here. 

For comparison, we also calculate the properties of accreted NS crusts making use of the experimental data from AME 1971
and the MB model, as in the earlier studies of  \cite{hz1990a, hz1990b,HZ2003,HZ2008}. However, our approach is slightly different 
from that followed in \cite{hz1990a, hz1990b,HZ2003,HZ2008}. To be consistent with the ETFSI calculations, the equilibrium configuration 
at each pressure $P$ and for given proton number $Z$ is now determined by minimising $g(\acell, Z, P)$ without imposing any constraint on the number 
of neutrons in the clusters or in the neutron gas (except of course the requirement of a fixed 
baryon number $\acell$ in the WS cell). In this way, neutron emissions and absorptions by clusters 
are taken into account. Because of the separate treatment of clusters and gas, the number $N$ of 
neutrons in clusters is constrained to take integer value.

\section{Structure and composition of accreted crust}
\label{sect:acc.crust.composition}

The internal constitution of the crust of a NS is of utmost importance for studying transport properties, which in turn can affect the cooling of the star, the evolution of the magnetic field or the seismic activity \citep{lrr}. Using the microscopic models described in Section \ref{sect:micro}, we have determined the composition of accreted NS crusts considering X-ray burst ashes made of $^{56}$Fe. 

\subsection{Outer crust}

\begin{figure}
\begin{center}
\includegraphics[width=\hsize]{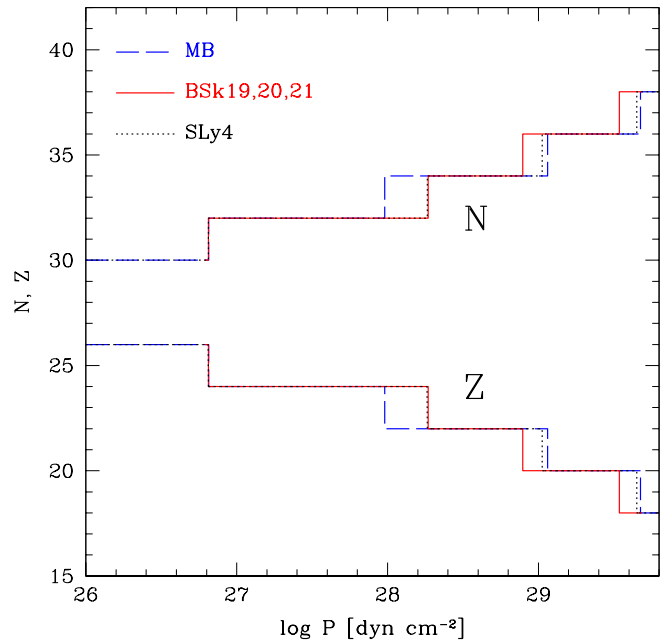}
\end{center}
\caption{Proton number $Z$ and neutron number $N$ of nuclei in the outer crust of an accreted NS for X-ray 
burst ashes made of 
$^{56}$Fe. Results are shown for the model of \cite{mb1977} (MB) and for HFB
nuclear mass models using different EDFs: BSk19, BSk20 and BSk21 from \cite{goriely2010} and SLy4 from \cite{chabanat1998}. 
}
\label{fig:acc-n-z-out}
\end{figure}

The variations of the proton number $Z$ and neutron number $N$ with increasing pressure are shown in Fig.~\ref{fig:acc-n-z-out}. The value of $Z$ systematically 
decreases with increasing depth due to electron captures, as discussed in Section \ref{sect:micro}. 
The HFB models with different EDFs and the MB model predict the same composition for the outer crust, but a different stratification. 
The first two shallowest layers are completely determined by measured masses, and the associated transition densities and pressures are therefore the same for all models. 
The boundaries of the densest layers are found to be model dependent due to the lack of experimental data. 

At high enough densities, electron captures by $^{56}$Ar are predicted by all models to emit free neutrons, thus marking the bottom of the outer crust. With the approach proposed by \cite{chamel2015}, the transition is found to be very smooth, in agreement with nuclear network calculations \citep{Gupta2008,LauBeard2018}. 
However, the neutron-drip density and pressure are found to depend on the details of the nuclear mass model employed, as previously analysed by \cite{fantina2016}. The properties of accreted NS crust at the neutron-drip transition are summarised in Table~\ref{tab:ndrip}\footnote{Values in Table~\ref{tab:ndrip} slightly differ from those in Table~V in \cite{chamel2015}; at variance with the latter work we now include electron exchange and polarisation corrections.}. The values for the neutron-drip density and pressure are found to differ  very little (less than 1\%) from those obtained in a strictly mean-nucleus treatment except for the model of MB, as previously shown by \cite{chamel2015}. Very accurate analytical formulas for the pressure at the transition between two adjacent strata as well as the average baryon number densities of each layer can be found in \cite{chamel2016b}.

\subsection{Inner crust}
\label{subsect:composition.icrust}

\begin{figure}
\begin{center}
\includegraphics[width=\hsize]{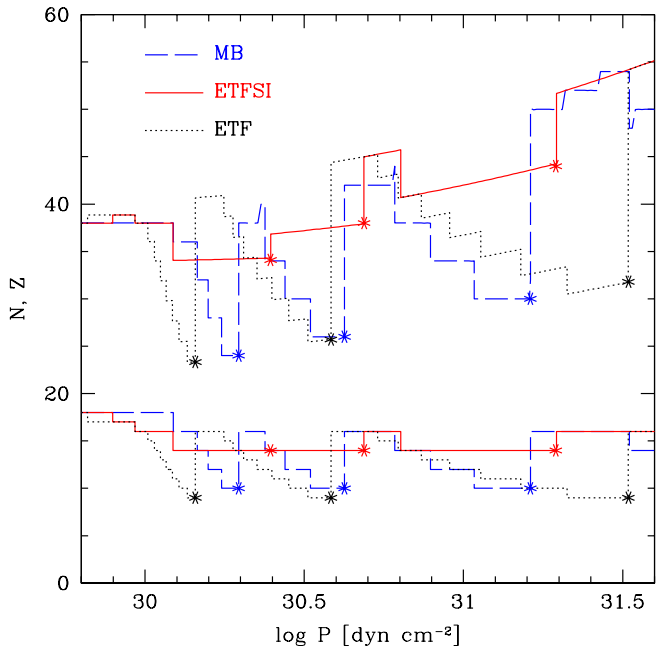}
\end{center}
\caption{
Proton number $Z$ and neutron number $N$ of nuclear clusters in the inner crust of an accreted NS for X-ray burst ashes made of 
$^{56}$Fe. Pycnonuclear reactions are marked by asterisks. Results are shown for the model of \cite{mb1977} (MB), and for the model based on the BSk19 EDF comparing two different treatments: ETFSI vs ETF (no shell correction)
}
\label{fig:acc-n-z}
\end{figure}

\begin{figure}
\begin{center}
\includegraphics[width=\hsize]{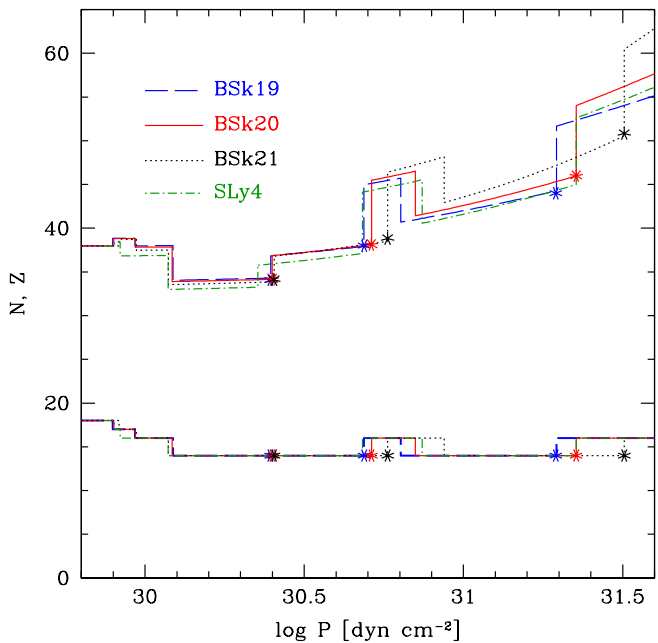}
\end{center}
\caption{
Proton number $Z$ and neutron number $N$ of nuclear clusters in the inner crust of an accreted NS for X-ray burst ashes made of 
$^{56}$Fe. Pycnonuclear reactions are marked by asterisks. Results are shown for the ETFSI models based on the EDFs BSk19, BSk20, BSk21, and SLy4. 
}
\label{fig:acc-n-z-bsk}
\end{figure}

The composition obtained with the EDF theory is substantially different from that predicted by the MB model in the inner part of the crust, as shown in Fig.~\ref{fig:acc-n-z}. In particular, the ETFSI model with the BSk19 EDF predicts that a large part of the inner crust 
is made of $Z=14$ clusters, whereas the MB model leads to a highly stratified crust with a high number of layers each 
of which contains a different type of cluster with $Z$ 
ranging from 10 to 18. Similar differences can be seen on the variations of the neutron number. In the EDF theory, nuclear 
clusters are thus found to be more stable against electron captures and pycnonuclear reactions than in the MB model. 
In the absence of electron captures and pycnonuclear reactions, the compression of a matter element is accompanied by a 
slight increase of the neutron number caused by the external neutron gas pressure which ``pushes'' back  neutrons into the clusters. 
The discrepancies between the EDF theory and the MB model can be attributed to the inclusion of nuclear shell effects in the EDF theory. In particular, the occurrence of crustal regions made of silicon isotopes supports the predictions of \cite{dutta2004} that $Z=14$ is a magic proton number in a dense stellar environment. As a matter of fact, $^{34}$Si (with $Z=14$ and $N=20$) appears as a doubly magic nucleus \citep{baumann1989}, as well as $^{42}$Si (with $Z=14$ and $N=28$) \citep{fridmann2005}. It has been recently suggested that $^{48}$Si (with $Z=14$ and $N=34$) could also be doubly magic \citep{li2016}. The 
existence of the magic neutron number $N=34$ was experimentally confirmed,  especially from measurements of $^{54}$Ca \citep{steppenbeck2013}. Quite remarkably, our ETFSI calculations predict the existence of $^{48}$Si in some layers despite 
our neglect of neutron-shell corrections. The appearance of $N=34$ in the extremely neutron-rich environment of the inner crust 
of accreted NSs might be driven by nuclear symmetry effects, and the existence of the proton magic number $Z=14$. 
To better illustrate the role of nuclear shell effects, we have performed calculations in the ETF approximation with no shell corrections. The resulting composition is strikingly different from that obtained in the full ETFSI treatment, as shown in Fig.~\ref{fig:acc-n-z}. In the absence of nuclear shell effects, a compressed matter element undergoes successive electron captures until the proton number becomes low enough for pycnonuclear reactions to occur. As expected, the ETF approximation leads to similar variations of $Z$ with pressure as those found with the MB model. As shown in Fig.~\ref{fig:acc-n-z-bsk}, all four EDFs considered here predict similar compositions.

\subsection{Accreted vs catalysed crusts}
\label{subsect:composition.acc.vs.cat}

Although both accreted and catalysed crusts contain $^{56}$Fe in their outer part,  the composition of their inner layers are remarkably different. 
As discussed in Section~\ref{sect:micro}, the proton number $Z$ in accreted crust systematically decreases with increasing depth due to electron captures unless pycnonuclear reactions occur. But even in this case, the daughter nucleus is generally highly unstable against electron captures so that $Z$ remains lower than $26$ in all regions of the crust. In contrast, the catalysed crust was found to be made of much heavier elements with $Z\geq 26$ within the EDF theory \citep{pearson2011,pearson2012}.

\begin{table}
\centering
\caption{Neutron-drip transition in the crust of accreting NSs, as predicted by different nuclear mass models for $^{56}$Fe ashes: atomic number $Z$ of the dripping nucleus, number of
 emitted neutrons, mass-energy density and corresponding pressure. Values in parenthesis are results obtained using a strictly mean-nucleus treatment. }\smallskip
\label{tab:ndrip}
\begin{tabular}{ccccc}
\hline \noalign {\smallskip}
&$Z$ &$\Delta N$ &$\rho_{\rm drip}$  & $P_{\rm drip}$ \\
&&&{\tiny ($10^{11}$ g\,cm$^{-3}$)} & {\tiny ($10^{29}$ dyn\,cm$^{-2}$)}\\
\hline \noalign {\smallskip}
 HFB-19 & 18&1&4.48 (4.50) &9.02 (9.06)\\
 HFB-20 & 18&1& 4.50 (4.52) &9.06 (9.11)\\
 HFB-21 & 18&1& 4.38 (4.40) &8.75 (8.79)\\
 HFB SLy4   & 18&1& 4.58 (4.60) &9.29 (9.33) \\
 MB     & 18&1& 5.09 (5.65) &  10.7 (12.3) \\
\end{tabular}
\end{table}

\begin{table}
\centering
\caption{Neutron-drip transition in the crust of catalysed NSs, as predicted by different nuclear mass models: atomic number $Z$ and neutron number $N$ of the dripping nucleus, mass-energy density and corresponding pressure. 
}\smallskip
\label{tab:ndrip-cat}
\begin{tabular}{ccccc}
\hline \noalign {\smallskip}
&$Z$ &$ N$ &$\rho_{\rm drip}$  & $P_{\rm drip}$ \\
&&&{\tiny ($10^{11}$ g\,cm$^{-3}$)} & {\tiny ($10^{29}$ dyn\,cm$^{-2}$)}\\
\hline \noalign {\smallskip}
HFB-19   & 38 & 88 &  $4.39$ & $7.89$ \\
HFB-20   & 38 & 88 &  $4.38$ & $7.87$ \\
HFB-21   & 38 & 86 & $4.29$  & $7.82$ \\
HFB SLy4     & 38 & 82&  $4.10$  & $7.68$ \\
MB       & 36 & 96 &  $2.49$ & $4.52$  \\
\hline
\end{tabular}
\end{table}

\begin{figure}
\begin{center}
\includegraphics[width=\hsize]{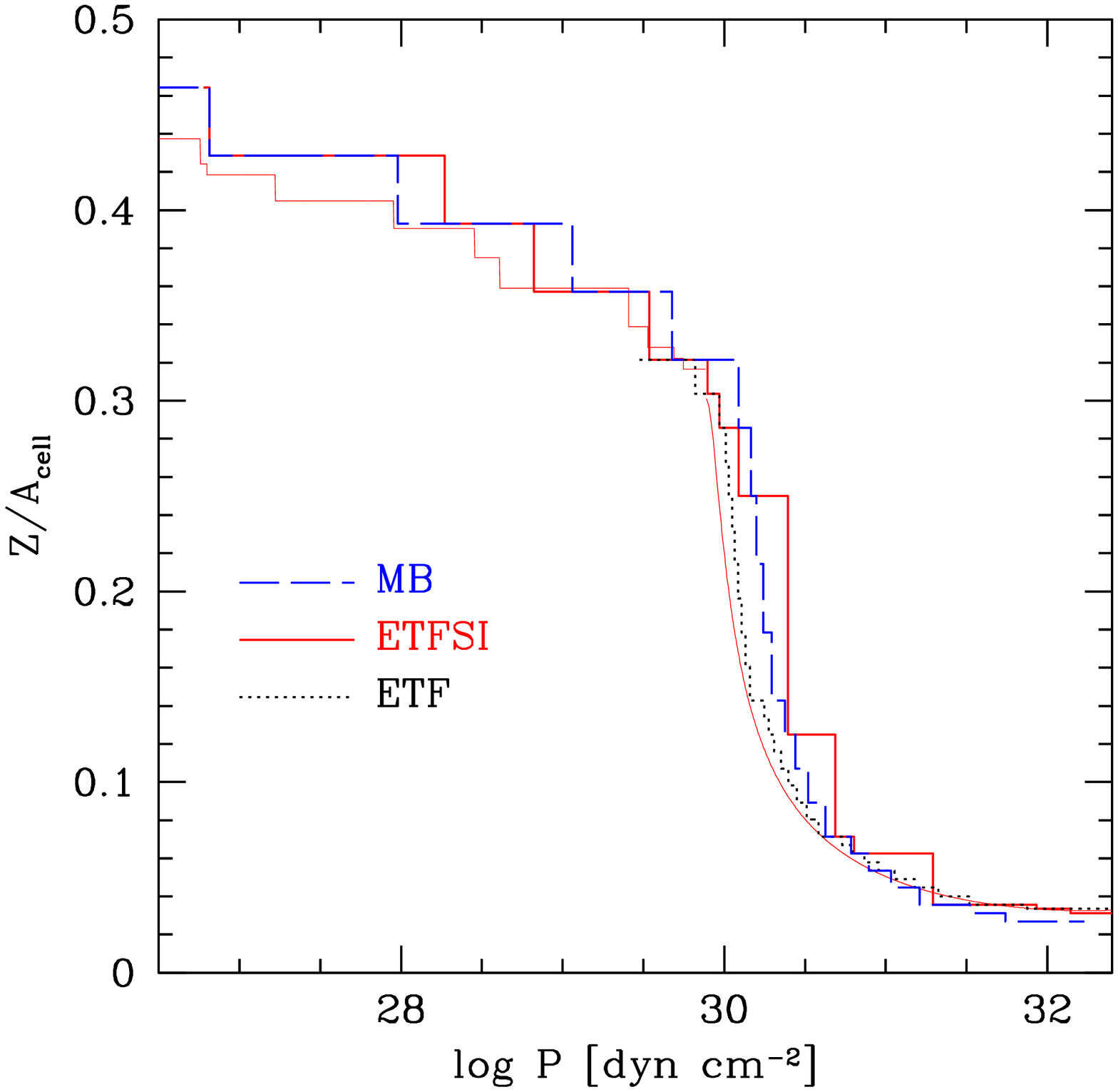}
\includegraphics[width=\hsize]{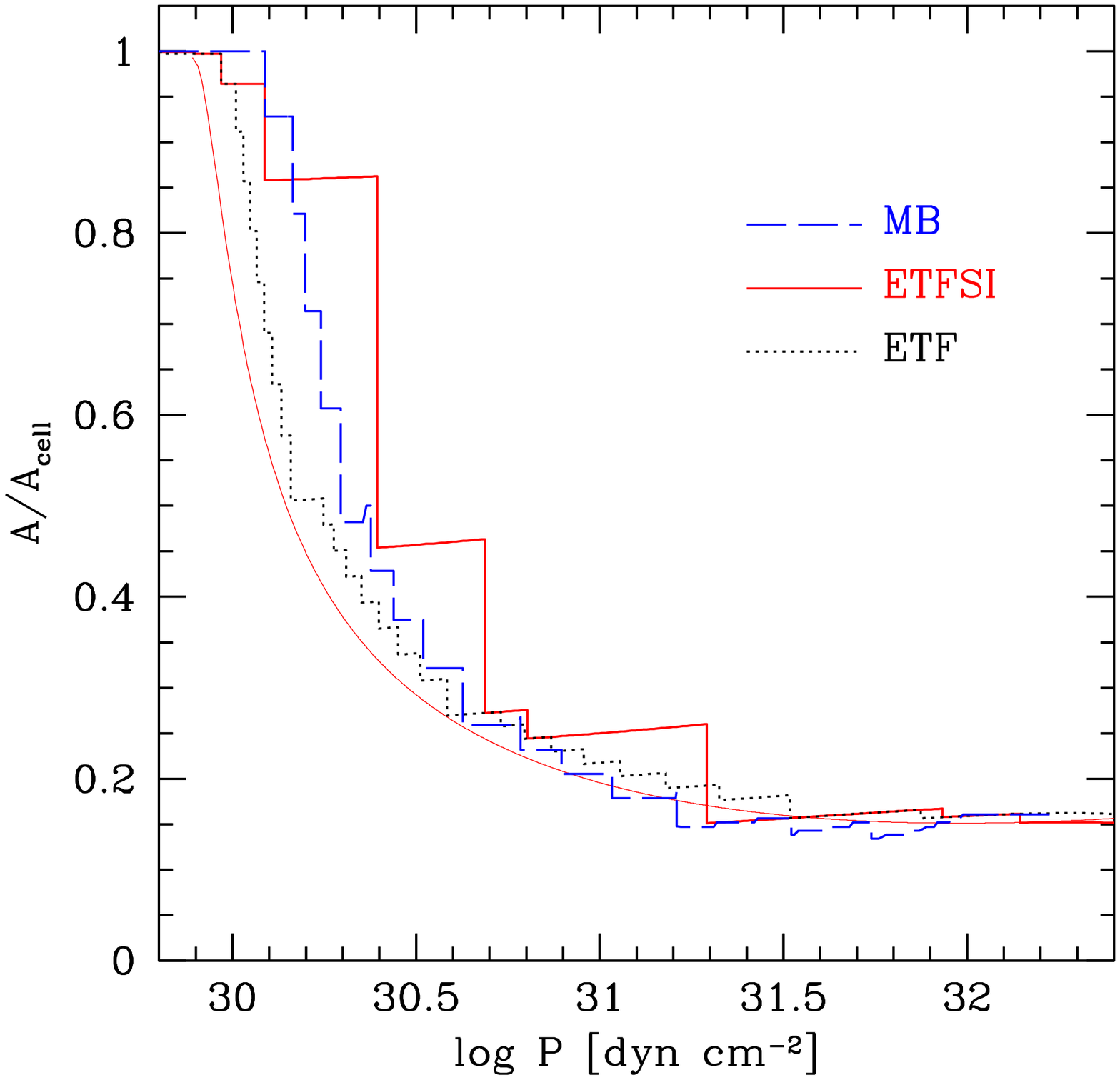}
\end{center}
\caption{Proton fraction (top panel) and fraction of nucleons in nuclei (bottom panel) for accreting NS crust and catalysed neutron star crust (for the model based on the BSk19 EDF, thin solid line). Results are shown for the model of \citet{mb1977} (MB), and for the model based on the BSk19 EDF comparing two different treatments for the inner crust: ETFSI vs ETF (no shell correction).}
\label{fig:yp-x}
\end{figure}

\begin{figure}
\begin{center}
\includegraphics[width=\hsize]{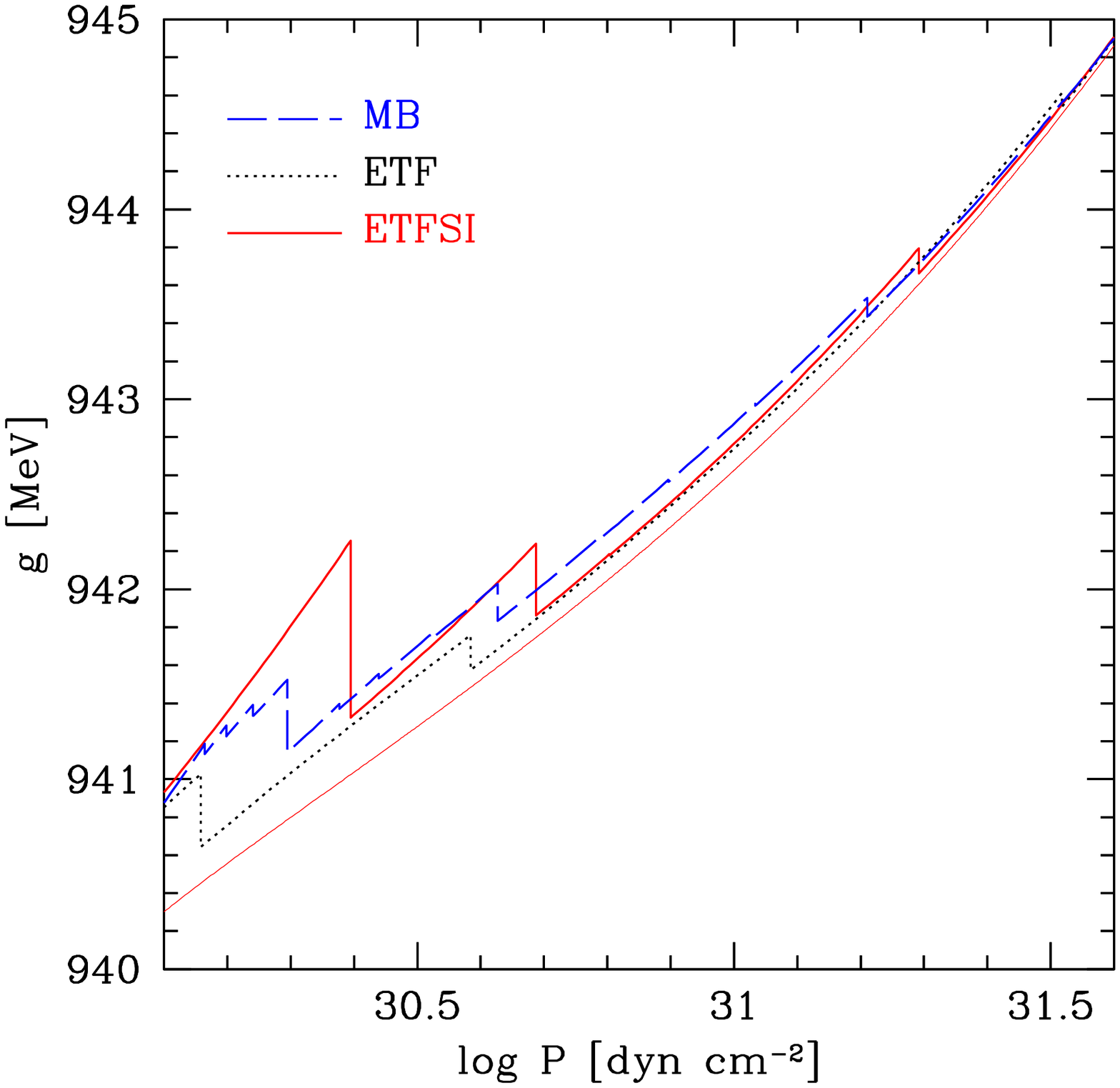}
\includegraphics[width=\hsize]{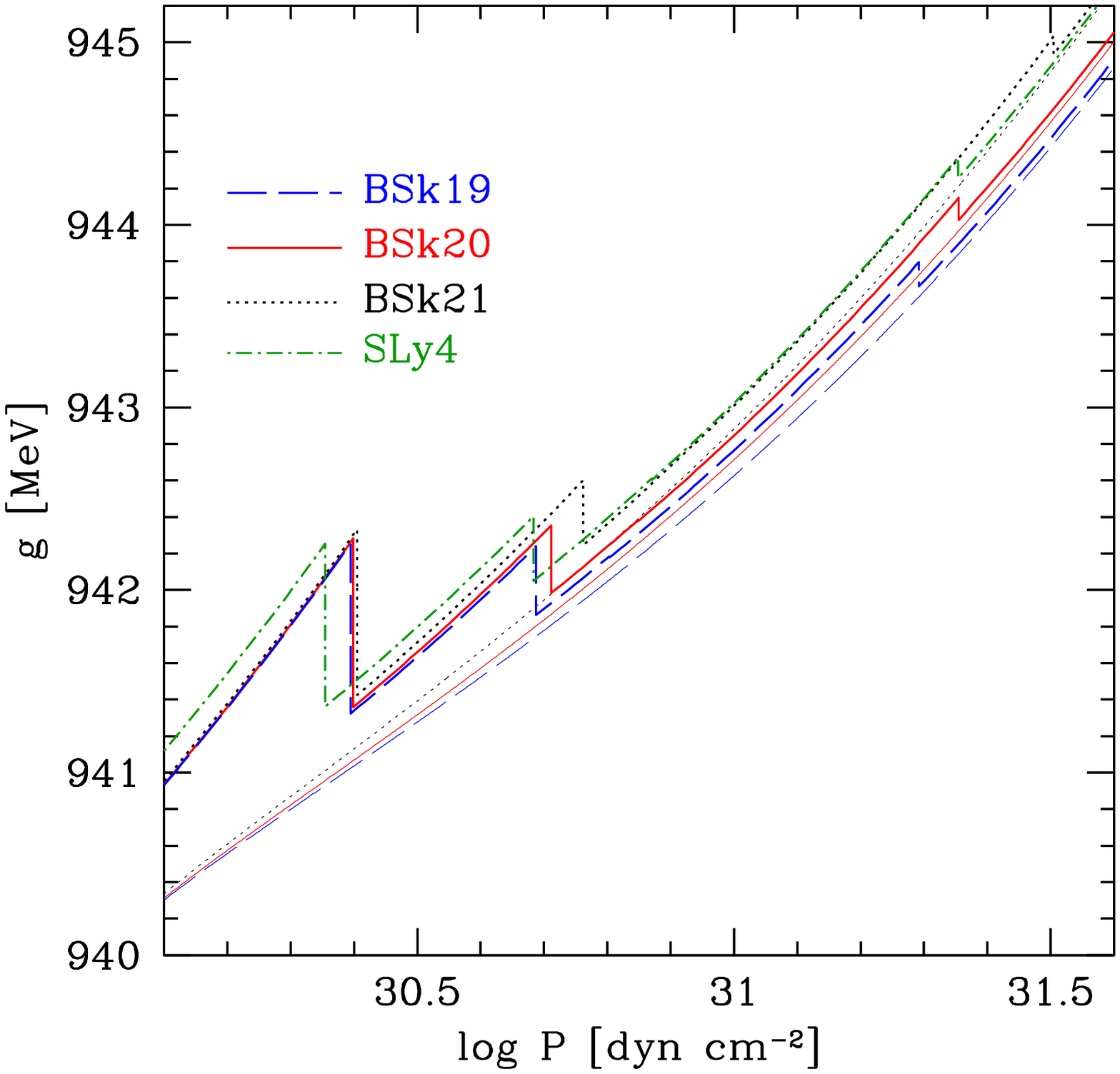}
\end{center}
\caption{Gibbs free energy per nucleon $g$ in accreting NS crust and catalysed NS crust (thin bottom lines). Top panel: predictions from the model of \cite{mb1977} (MB) and from the model based on the BSk19 EDF comparing two different treatments for the inner crust: ETFSI vs ETF (no shell correction). Bottom panel: predictions from models based on differents EDFs. See text for detail.
}
\label{fig:dmubdlogP}
\end{figure}

In spite of the radically different constitution of accreted and catalysed crusts, 
in both cases matter becomes progressively more neutron rich with increasing pressure due to gravitational settling (see, e.g. \citet{chamel2016b} for a generic argument based on Le Chatelier's principle; see also \citet{deblasio2000}). The properties of catalysed NS crust at the neutron-drip transition are summarized in Table~\ref{tab:ndrip-cat}, where we have listed, for the different models, the $Z$ and $N$ of the dripping nucleus, the mass-energy density, and the corresponding pressure.
For the EDF models HFB19, HFB-20, HFB-21, and SLy4, the latters have been calculated as described in \citet{chamel2016} by solving numerically their Eqs.~(13)-(14) and including the electron exchange and polarisation corrections.
As shown in Figs.\ref{fig:yp-x} and \ref{fig:dmubdlogP}, 
the proton fraction $Z/A_\textrm{cell}$, the fraction $A/A_\textrm{cell}$ of nucleons contained in nuclei, and the Gibbs free energy per nucleon  $g$ follow the same trend in both accreted and catalysed crusts. This can be understood as follows.
Although the values of $Z$ and $A$ arise from a detailed balance between surface and Coulomb effects (see, e.g. \citet{lrr}), the ratios $y_p\equiv Z/A_\textrm{cell}$ 
and $x\equiv A/A_\textrm{cell}$ can be roughly estimated considering the coexistence of two distinct homogeneous phases: 
a dense nuclear matter phase representing clusters (nuclei in the outer crust)  and a dilute neutron gas. Electrons are uniformly distributed, and therefore 
permeate the two phases. It should be remarked that the errors induced by this simple description decrease with increasing 
pressure as matter become progressively more homogeneous. 
The composition of catalysed crust at any given pressure $P$ is obtained by  minimising $g$. In accreted crusts, matter is not in full thermodynamic equilibrium. The composition is determined from a \emph{constrained} minimisation of $g$ with $A_\textrm{cell}$ fixed to the mass number of the ashes of X-ray bursts, and $Z$ can only be reduced unless pycnonuclear 
reactions are allowed. However, in the bulk approximation, the Gibbs free energy per nucleon does not depend on $Z$, $A$, and $A_\textrm{cell}$ separately but only on the ratios $Z/A_\textrm{cell}$ and $A/A_\textrm{cell}$, which can be treated as continuous variables at this level of approximation. Consequently, not only  $y_p(P)$ and $x(P)$, but also $g$ are the same for both accreted and catalysed crusts. 

The most remarkable differences between accreted and catalysed crusts concern the variation of the Gibbs free energy per nucleon with pressure, which is discontinuous in the former case and continuous in the latter case, as shown in Fig.~\ref{fig:dmubdlogP}. Jumps in $g$ correspond to the heat $Q_{\rm cell}$ released by electron captures and/or pycnonuclear reactions, see Eq.~(\ref{eq:heat-released}). In turn, $Q_{\rm cell}$ is very sensitive to the fine details of the nuclear structure, and therefore to the nuclear model employed. In particular, $Q_{\rm cell}$ is determined to a large extent by nuclear shell effects, as can be seen in Fig.~\ref{fig:dmubdlogP} from the comparison of the ETFSI and ETF results. 
By definition, no heat can be further extracted from catalysed matter. Therefore, $g$ increases continuously with pressure in catalysed crust, and is lower than the values obtained for accreted crusts. However, the differences become negligibly small at high pressures, as shown in Fig.~\ref{fig:dmubdlogP}.

\vskip 2mm

\section{Crustal heating}
\label{sect:acc.crust.heating}

\subsection{Outer crust}
\label{subsect:acc.ocrust.heating}

\begin{figure}
\begin{center}
\includegraphics[width=\hsize]{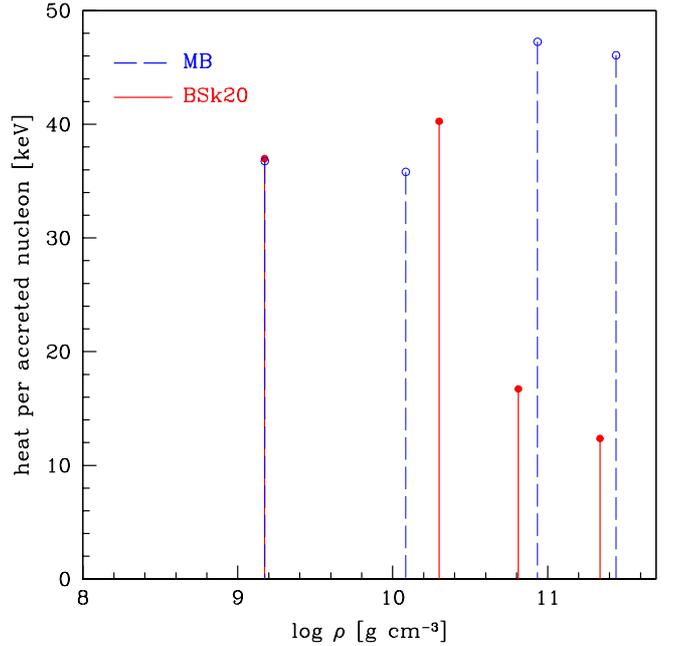}
\end{center}
\caption{Heat sources in the outer crust of accreting NS. Comparison between the model of \cite{mb1977} (MB) and the HFB mass model using the BSk20 EDF.}
\label{fig:e_sources_o1}
\end{figure}

\begin{figure}
\begin{center}
\includegraphics[width=\hsize]{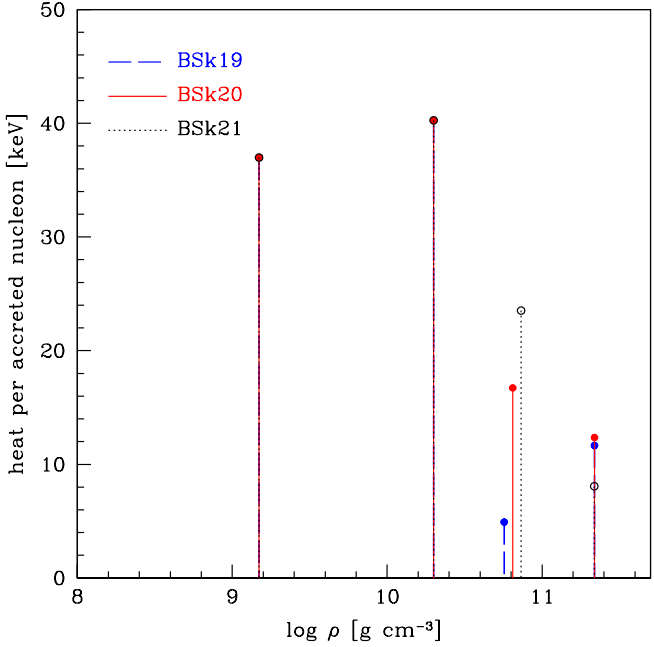}
\end{center}
\caption{Heat sources in the outer crust of accreting NS. Comparison between the HFB mass models using the BSk19, BSk20, and BSk21 EDFs. The first two reactions are fully determined by experimental measurements, and are the same for all HFB models. }
\label{fig:e_sources_o2}
\end{figure}

The heat sources, as obtained from different nuclear mass models, are indicated in Figs.~\ref{fig:e_sources_o1} and \ref{fig:e_sources_o2}. In the shallowest two layers of the outer crust, all models yield similar predictions since electrons are captured by nuclei whose masses are experimentally known. On the other hand, large discrepancies can be observed at higher densities between the different models. As discussed in Sect.~\ref{subsect:EDF}, the EDFs BSk19, BSk20, and BSk21 were all fitted to the atomic mass data from AME 2003~\citep{audi2003}, whereas SLy4 was fitted to a few atomic masses from AME 1995. To assess the reliability of the different models, we have compared  their predictions for the atomic mass excess of $^{56}$Sc, which has been only recently measured. As can be seen in Table~\ref{tab:masses-scandium}, the largest deviation between experimental and theoretical values is found for SLy4. The Brussels-Montreal EDFs considered here yield predictions closer to the experimental value, the best (worse) agreement being achieved with HFB-21 (HFB-19) consistently with the previous analysis of \cite{chamel2011}. To evaluate the impact of theoretical nuclear masses on the heat released, we have repeated the outer-crust calculations considering data from the older  AME 2003~\citep{audi2003} and AME 2012~\citep{audi2012}. As shown in Table \ref{tab:total.heat.ocrust}, the most robust predictions are given by HFB-21. On the contrary, the heat obtained with HFB SLy4 is increased by about a factor of two including the newly measured mass of $^{56}$Sc. For this model, the transitions from $Z=22$ to $Z=21$, and from $Z=21$ to $Z=20$ both occur in quasiequilibrium with no heat released.

\begin{table}
\centering
\caption{Experimental atomic mass excess (in MeV) of $^{56}$Sc from AME 2016~\citep{audi2016}, as compared to the prediction from different models.}
\smallskip
\label{tab:masses-scandium}
\begin{tabular}{ccccc}
\hline \noalign {\smallskip}
 AME 2016 &  HFB SLy4 & HFB-19  & HFB-20 & HFB-21 \\
 \hline \noalign {\smallskip}
  -24.85  & -28.22   & -26.40 & -25.82 & -25.23 \\
\hline
\end{tabular}
\end{table}

\begin{table}
\centering
\caption{Total heat per accreted nucleon (in MeV) deposited in the outer crust of an accreting NS, for X-ray burst ashes made of $^{56}$Fe using experimental atomic masses supplemented with theoretical predictions from HFB  models based on different EDFs (BSk19-21, SLy4). For comparison, results are shown using data from AME 2016~\citep{audi2016}, AME 2012~\citep{audi2012}, and AME 2003~ \citep{audi2003}. See text for details.}
\smallskip
\label{tab:total.heat.ocrust}
\begin{tabular}{cccc}
\hline \noalign {\smallskip}
          & AME 2003  & AME 2012  & AME 2016 \\
          \hline \noalign {\smallskip}
HFB SLy4  & 0.0872    & 0.0904    & 0.2030   \\
HFB-19    & 0.0858    & 0.0938    & 0.1522   \\ 
HFB-20    & 0.0983    & 0.1231    & 0.1440   \\
HFB-21    & 0.1008    & 0.1089    & 0.1255   \\
\hline
\end{tabular}
\end{table}

\begin{figure}
\begin{center}
\includegraphics[width=\hsize]{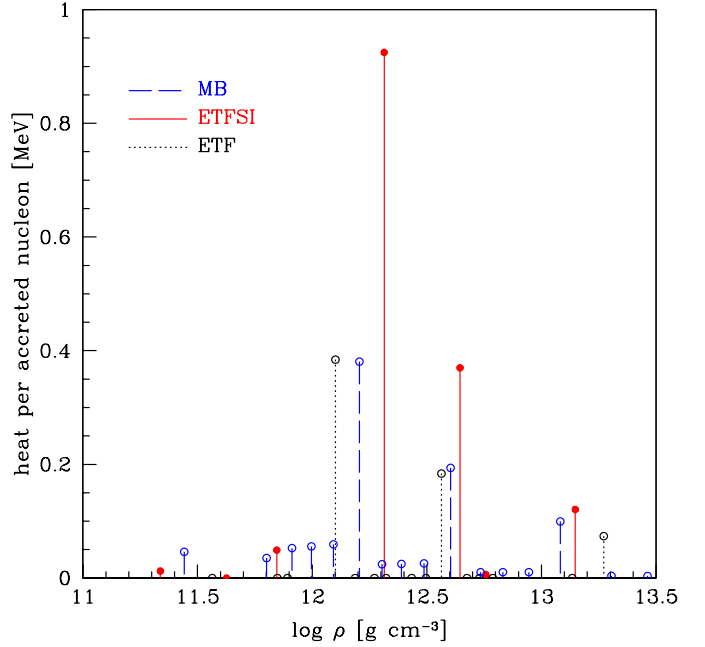}
\end{center}
\caption{Heat sources in the inner crust of accreting NS. Comparison between the model of \cite{mb1977} (MB) and the EDF theory using BSk19 for two different treatments: ETFSI and ETF. 
Pycnonuclear reactions correspond to the three largest peaks for each model. See text for detail.
}
\label{fig:e_sources_i1}
\end{figure}

\begin{figure}
\begin{center}
\includegraphics[width=\hsize]{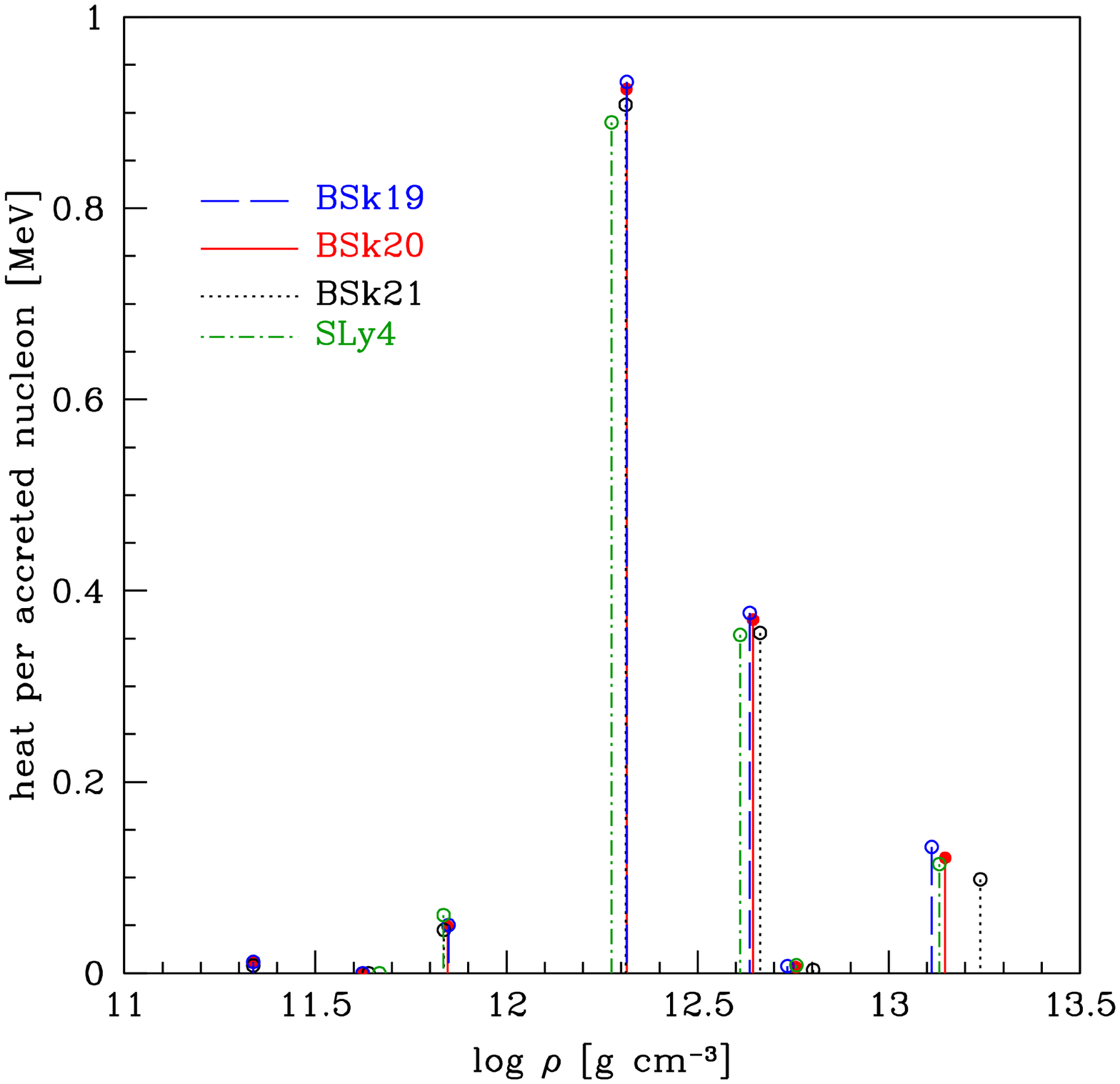}
\end{center}
\caption{Heat sources in the inner crust of accreting NS. Comparison between the ETFSI models based on the EDFs BSk19, BSk20,  BSk21 and SLy4. See text for detail.
}
\label{fig:e_sources_i2}
\end{figure}

As summarised in Table~\ref{tab:total.heat}, the MB model yields intermediate values for the total heat. The most remarkable difference between the HFB and the MB models lies in the individual heat sources, which are very sensitive to the nuclear structure. Indeed, approximating the Gibbs free energy per nucleon by \citep{bps1971}

\begin{equation}
g(A,Z,P)\approx \frac{M^\prime(A,Z)c^2}{A}+\frac{Z}{A}\left(\mu_e -m_e c^2\right)\, ,
\end{equation}
the heat released by the capture of two electrons by the nuclei $(A,Z)$ is given by 
\begin{equation}\label{eq:heat-ocrust}
Q_\textrm{cell} \approx  -M^\prime(A,Z) c^2 - M^\prime(A,Z-2)c^2 + 2 M^\prime(A,Z-1)c^2 \, .
\end{equation}
From this equation, $Q_\textrm{cell}$ is expected to be comparatively smaller for nuclei near shell closure. This is confirmed by the HFB models with the accurately calibrated Brussels-Montreal EDFs: they predict a smaller heat released in the densest region of the outer crust, where electron captures involve nuclei with proton magic number $Z=20$. On the contrary, the HFB model with SLy4 leads to the largest heat source for the transition from $Z=22$ to $Z=20$, thus reflecting an improper description of the nuclear shell closure. 

In contrast to HFB models, the heat sources obtained from the MB model are roughly the same for all reactions. The MB model relies on the liquid-drop picture of the nucleus ignoring the underlying nuclear shell structure. It is based on a refined version of the semi-empirical mass formula of \cite{weis1935,bethe1936} 
\begin{equation}\label{eq:binding}
M^\prime(A,Z)c^2=Z(m_p+m_e)c^2+(A-Z) m_n c^2 - B(A,Z)\, ,
\end{equation}
where the binding energy $B(A,Z)$ is expressed as 
\begin{eqnarray}\label{eq:BW}
B(A,Z)&=&a_V A - a_S A^{2/3} - a_C \frac{Z^2}{A^{1/3}}- a_A\frac{(A-2Z)^2}{A}\nonumber \\ 
&& +\frac{(-1)^Z}{2}\biggl[1+(-1)^A\biggr]\frac{a_P}{\sqrt{A}}\, .
\end{eqnarray}
The first term in Eq.(\ref{eq:BW}) accounts for the bulk energy of symmetric nuclear matter, the second for the surface energy of the nucleus, the third for the Coulomb energy, the fourth for the symmetry energy, and the last for the ``pairing" energy. 
In the MB model, the coefficients are not constant but depend on the mean neutron and proton densities inside the nucleus. Ignoring these refinements here, using Eqs.~(\ref{eq:heat-ocrust}) and (\ref{eq:BW}), we find that the heat released during the capture of two electrons by nuclei $(A,Z)$ is independent of $Z$, and is approximately given by 
\begin{equation}\label{eq:heat-ocrust-BW}
Q_\textrm{cell} \approx \frac{4 a_P}{\sqrt{A}} - \frac{8 a_A}{A}  - \frac{2 a_C}{A^{1/3}}\, . 
\end{equation}
In reality, the heat released is weakly dependent on $Z$ because of the interactions between electrons and ions that we have neglected in Eq.(\ref{eq:heat-ocrust}). 
This analysis shows that the heat released is independent of volume and surface 
 terms in the mass formula (\ref{eq:BW}), but mainly arises from ``pairing''.  However, this contribution is partially cancelled by the symmetry energy term, and to a lesser extent by the  Coulomb term. Indeed, it is energetically favorable for a nucleus with an odd number of protons to transform into  a nucleus with an even number of protons because of pairing. On the other hand, this can only be realised through the capture of  an electron. However this process, which implies the conversion of the unpaired proton into a neutron, costs symmetry energy thus 
 reducing the heat released. In other words, high values of the symmetry energy coefficient $a_A$ would lead to the release of a small quantity of heat, and vice versa. Using values of the coefficients from the MB model with $A=56$, we find that the total heat released by the four electron captures in the outer crust is about 0.11 MeV. This value differs from that indicated in Table~\ref{tab:total.heat} due to the inclusion of density-dependent coefficients.

\begin{figure}
\begin{center}
\includegraphics[width=\hsize]{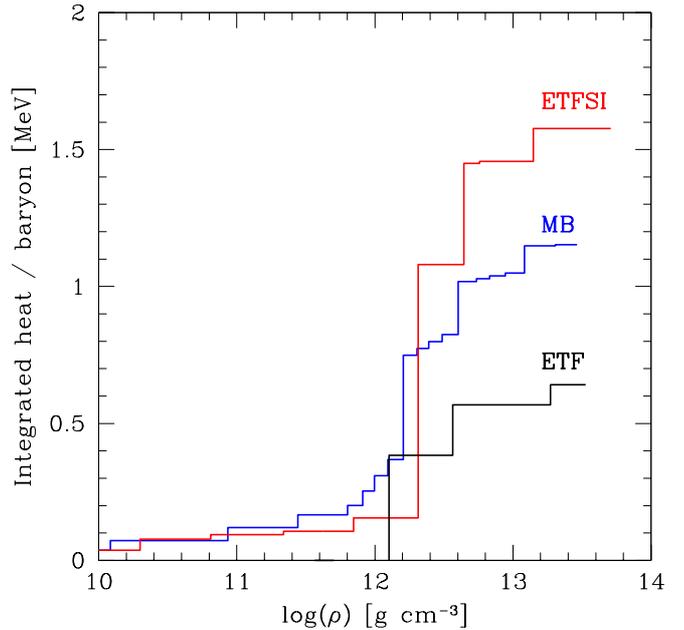}
\end{center}
\caption{Integrated heat in the crust of accreting NS. Comparison between the model of \cite{mb1977} (MB) and the EDF theory using BSk19 for two different treatments: ETFSI and ETF. See text for detail.}
\label{fig:qintegrated1}
\end{figure}

\begin{figure}
\begin{center}
\includegraphics[width=\hsize]{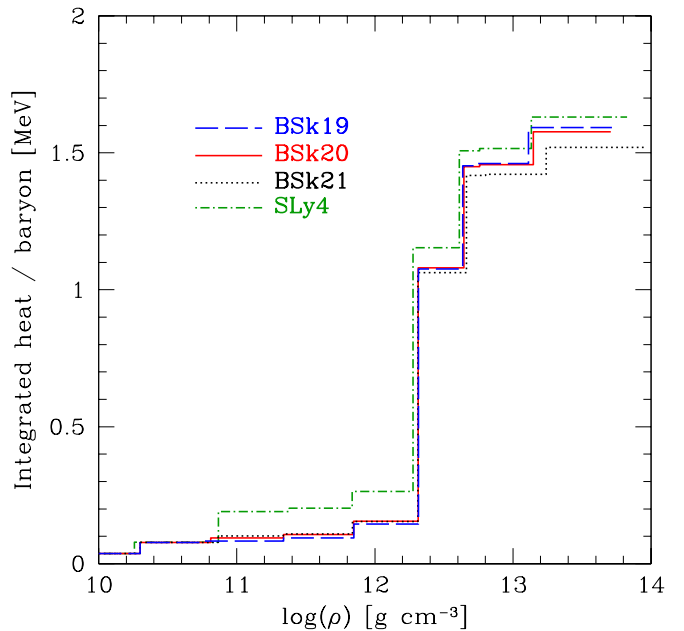}
\end{center}
\caption{Integrated heat in the crust of accreting NS. Comparison between the ETFSI models based on the EDFs SLy4, BSk19, BSk20, and BSk21. See text for detail.
}
\label{fig:qintegrated2}
\end{figure}

\subsection{Inner crust}
\label{subsect:acc.icrust.heating}

The heat sources in the inner crust of accreting NS, as predicted by different models, are indicated in Figs.~\ref{fig:e_sources_i1} and \ref{fig:e_sources_i2}. 
Most of the heat is released through electron captures as well as neutron emissions and absorptions  associated with pycnonuclear reactions. The number of pycnonuclear reactions is the same (3) for all models in the range
of densities shown in Figs.~\ref{fig:e_sources_i1} and \ref{fig:e_sources_i2}. 
However, the amount of heat released by these processes is found to be twice as small for the MB model ($\sim 0.67$ MeV) as compared to the ETFSI models ($\sim 1.4$ MeV). These discrepancies can be explained by the inclusion of nuclear shell effects in the ETFSI approach. Indeed, the ETF approximation leads to comparable heat release from pycnonuclear reactions ($\sim 0.64$ MeV) as the MB model. Without any nuclear shell corrections, pycnonuclear reactions are actually the only source of heat since electron captures and neutron emissions occur in quasi-equilibrium.

The integrated heat, defined by 
\begin{equation}
Q(P) = \sum_{j(P_j<P)}Q_j\, ,
\end{equation}
is presented in Figs.\ref{fig:qintegrated1} and \ref{fig:qintegrated2}. The total heat deposited in the crust is indicated in Table~\ref {tab:total.heat}. For the ETFSI models, the results are  quite similar for all four EDFs. In particular, the differences between SLy4 and BSk19-21 are less pronounced than  in the outer crust. Most of the heat is released from pycnonuclear reactions involving transitions from $Z=16$ to $Z=14$. The close agreement between the different EDFs reflects the fact that the nuclear closure at $Z=14$ is a robust prediction from HFB calculations. 

The total heat obtained for SLy4 is significantly lower than the value of 2.5 MeV found by \cite{steiner2012} for the same EDF within a liquid-drop picture and considering pure ashes made of $^{56}$Ni. The discrepancy cannot be simply explained by the conversion of $^{56}$Ni into $^{56}$Fe, since the associated heat released is estimated as 0.04 MeV from Eq.~(\ref{eq:heat-ocrust}) using the data of \cite{audi2003}. 
The large heat predicted by \cite{steiner2012} must thus lie in the different treatment of crustal matter. In particular, the SLy4 EDF enters into the calculations of \cite{steiner2012} only through the bulk nuclear energy of matter. The surface and Coulomb energies were calculated phenomenologically. More importantly, the  added empirical corrections for pairing and shell effects fail to reproduce some magic numbers, especially for very neutron-rich nuclei, such as the neutron numbers $N=20$ and $N=34$ (see also the discussion in Section~\ref{subsect:composition.icrust}). 
However, the total heat deposited in the crust is very sensitive to nuclear shell effects. As illustrated in Fig.~\ref{fig:qintegrated1}, the heat is indeed considerably reduced, by about a factor of 3, in pure ETF calculations without any shell correction. The MB model yields an  intermediate value of $1.16$~MeV due to the inclusion of pairing effects. This amount of the total deposited heat is substantially lower than the value of $1.85$~MeV found by \cite{HZ2008}. The discrepancy arises from the more realistic treatment of the neutron emissions and absorptions by clusters in this work, as discussed in Section~\ref{sect:inner.crust.model}. In particular, in previous works of \cite{hz1990a,hz1990b,HZ2003,HZ2008} neutrons produced by electron captures were confined in the clusters, thus introducing artificial shell effects. The role of nuclear shell effects is more closely analysed in the next section. 

\begin{table}
\centering
\caption{Total heat per accreted nucleon (in MeV) deposited in the outer and inner crust of an accreting NS, for 
X-ray burst ashes made of $^{56}$Fe for different nuclear models: microscopic 
 models based on different EDFs (BSk19-21, SLy4) compared to the liquid drop model of Mackie\&Baym (MB). See text for details.}
\smallskip
\label{tab:total.heat}
\begin{tabular}{cccc}
\hline \noalign {\smallskip}
 &  outer crust  & inner crust &  total heat \\
 \hline \noalign {\smallskip}
  BSk19 &  0.152  &  1.499  &  1.651    \\
  BSk20 & 0.144   & 1.471     &1.615     \\
  BSk21 &   0.125   &1.410   & 1.535   \\
  SLy4 & 0.203 &1.456 & 1.659 \\
  MB  & 0.166 & 0.976 & 1.142 \\
\hline
\end{tabular}
\end{table}

\subsection{Importance of nuclear shell effects}

\begin{figure}
\begin{center}
\includegraphics[width=.8\hsize]{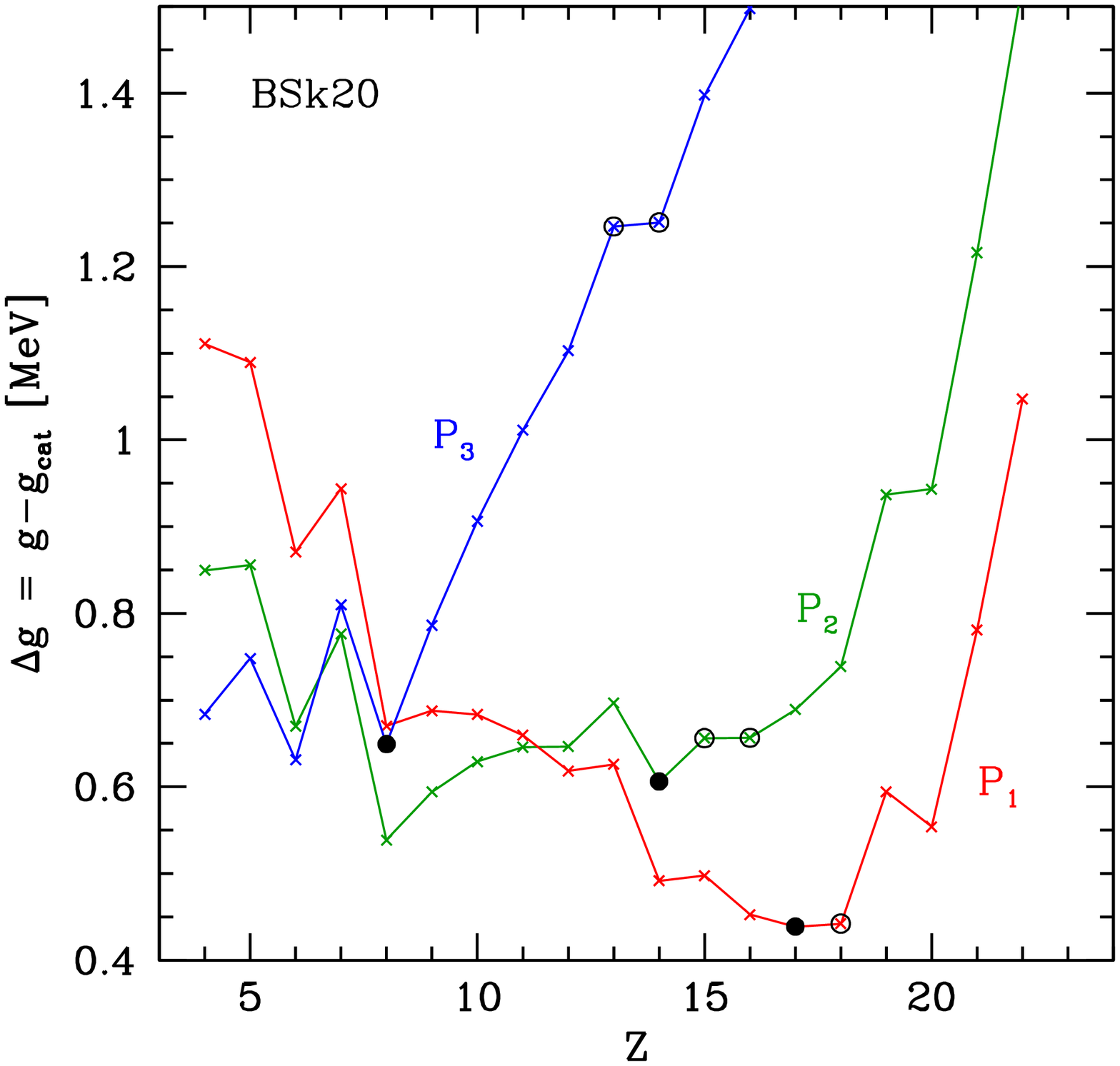}
\includegraphics[width=.8\hsize]{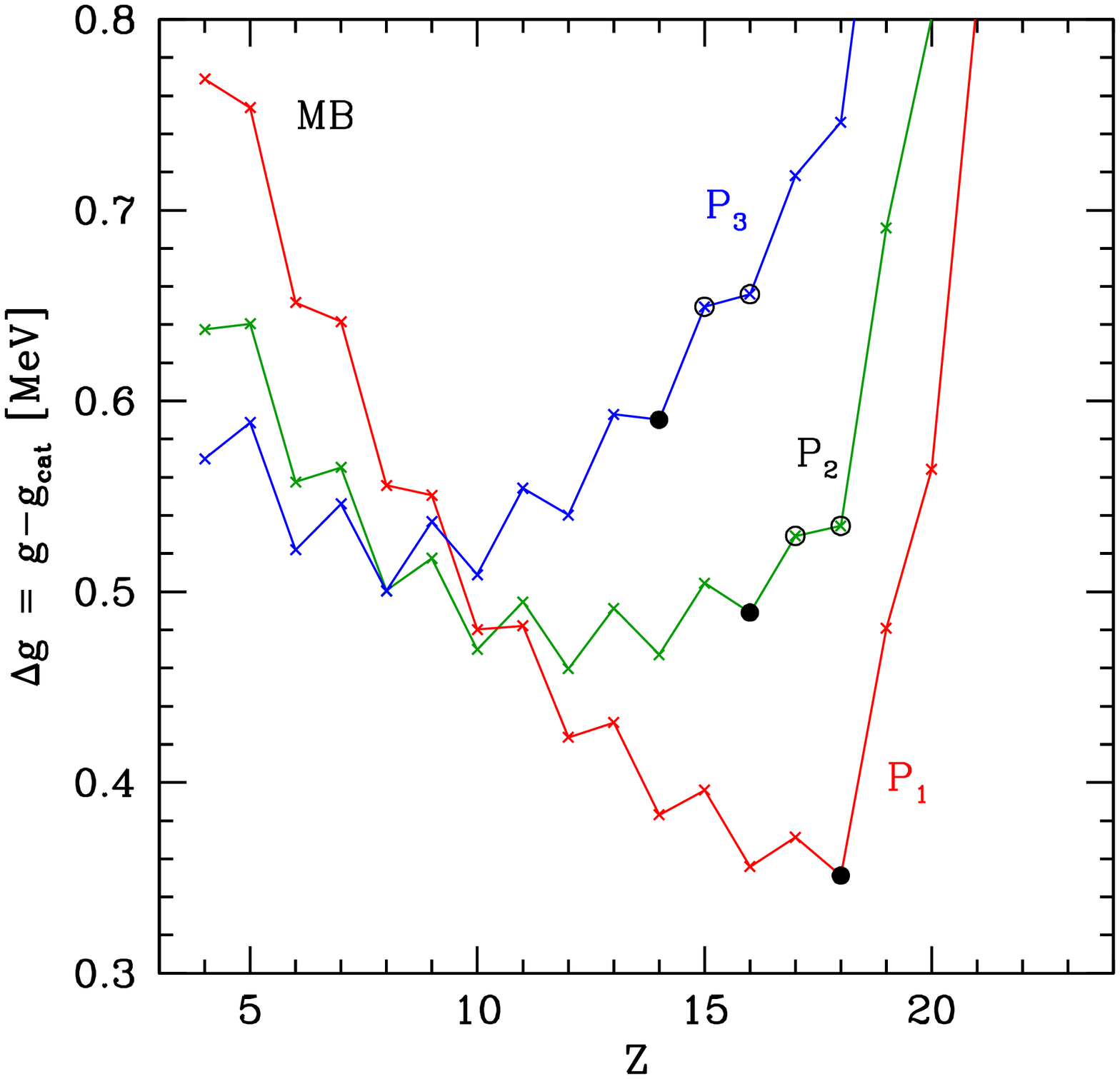}
\end{center}
\caption{Gibbs free energy per nucleon for accreted NS crust (renormalized to that of catalysed matter) for nuclear clusters with different $Z$, as predicted by the model of \cite{mb1977} (lower panel) and the ETFSI model with the BSk20 EDF (upper panel). With increasing pressure ($P_1<P_2<P_3$), the energetically preferred value for $Z$ decreases. For BSk20 
$(P_1, P_2, P_3)=(0.82, 1.22, 2.6)\cdot10^{30}$ erg$\,$cm$^{-3}$, for MB (bottom panel) 
$(P_1, P_2, P_3)=(1.03, 1.28, 1.54)\cdot10^{30}$ erg$\,$cm$^{-3}$. Solid dots represent stable clusters at given pressure whereas circles denote the parent clusters for the last reactions just before the given pressure is reached.
}
\label{fig:dmuZ}
\end{figure}

The importance of nuclear shell effects for the evolution of an accreted matter element and deep crustal heating is illustrated in Fig.~\ref{fig:dmuZ}. The variation of the Gibbs free energy per nucleon as a function of $Z$ is shown for three different values of the pressure $P_1<P_2<P_3$ for the ETFSI model based on the EDF BSk20 (the other EDFs yielding qualitatively similar results) and for the MB model.

At pressures $P\lesssim P_1=8.2\times 10^{29}$~erg~cm$^{-3}$, the accreted material obtained from the ETFSI model consists of $^{56}$Ar ($Z=18$), corresponding to the global minimum of $g$. At $P=P_1$, the minimum of $g$ shifts to $Z=17$: $^{56}$Ar nuclei capture one electron and transform into $^{56}$Cl. A similar situation occurs as the pressure attains $P=9.3\times 10^{29}$~erg~cm$^{-3}$, $^{56}$Cl being converted into $^{56}$S. With further compression, $^{56}$S nuclei become unstable against electron captures at pressure $P\simeq P_2=1.22\times 10^{30}$~erg~cm$^{-3}$ (the local minimum of $g$ at $Z=16$ disappears) and the matter element evolves to the local minimum of $g$ corresponding
to $Z=14$ with the release of heat equal to the relative depth of this minimum. 
Since $Z=14$ is a magic number, this local minimum persists over a large range of pressures, up to $P_3=2.6\times 10^{30}$~erg~cm$^{-3}$. As a consequence, nuclei $^{56}$Si undergo a chain of six electron captures accompanied by neutron emissions.
These reactions are highly exothermic. 

The evolution of an accreted matter element is radically different for the MB model.  
With the inclusion of pairing effects, the local minima of $g$ correspond to even values of $Z$. As a consequence, the matter element undergo many more electron captures. For instance, $^{56}$Ar nuclei present at pressure $P_1=1.03\times 10^{30}$~erg~cm$^{-3}$ become unstable at pressure $P_2=1.28\times 10^{30}$~erg~cm$^{-3}$, and transform into $^{56}$S. Similarly, $^{56}$S nuclei are converted into $^{56}$Si at pressure $P_3=1.54\times 10^{30}$~erg~cm$^{-3}$. The changes of pressure between two successive reactions are much smaller for the MB model than in the ETFSI approach. In the example discussed above, the pressure difference $P_3-P_2$ is about five times smaller in the MB model than in the ETFSI model. This means that in the ETFSI model five times larger amount of mass  (or five times longer time for given accretion rate) is needed for a matter element to be compressed to the density at which the third reaction takes place. On the other hand, the heat released is larger 
by a similar factor (5-6) so that the total heat deposited in the crust is of the same order as predicted by the MB model.

\begin{figure}
\begin{center}
\includegraphics[width=\hsize]{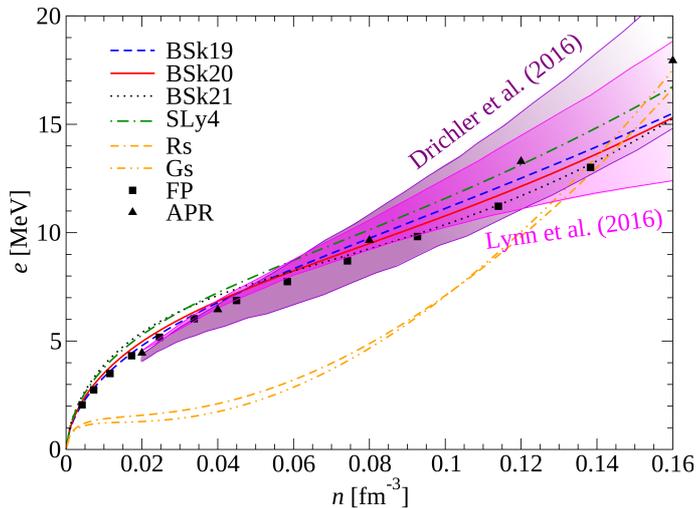}
\end{center}
\caption{Equation of state of pure neutron matter, as predicted by different EDFs (lines). For comparison, the microscopic calculations of \cite{fp1981} (FP) and \cite{apr1998} (APR) based on realistic nucleon-nucleon interactions are shown by symbols. The shaded areas are the constraints inferred from the ab-initio calculations of \cite{drichler2016} and  \cite{lynn2016}. 
} 
\label{fig:PNM}
\end{figure}

\subsection{Importance of realistic neutron-matter equation of state}
\label{subsect:EOS-PNM}

As can be seen in Fig.~\ref{fig:e_sources_i2}, most of the heat is deposited in the intermediate regions of the crust from pycnonuclear reactions followed by a chain of 
electron captures, neutron emissions and absorptions. In turn, the presence of free neutrons modify the properties of the clusters, and thereby their stability against further electron captures. As a matter of fact, the surrounding neutrons are crucial for the stability of the clusters, which would otherwise disintegrate. The neutron-matter stiffness was shown to have a strong influence on the composition of the inner crust of a nonaccreted NS \citep{goriely2005}, and it is expected to play a similarly important role for accreted NSs. All the EDFs considered in this work are consistent with realistic neutron-matter EoSs, as shown in Fig.~\ref{fig:PNM}. On the contrary, the Gs and Rs EDFs \citep{fried86} considered by \cite{steiner2012} are incompatible with microscopic calculations at the low densities relevant for NS crusts. This may explain the large amounts of heat found with these EDFs, namely 4.3 MeV/nucleon and 4.8 MeV/nucleon respectively. 

\section{Discussion and conclusions}
\label{sect:discussion}

We have studied heating in the crust of accreting NSs following a more realistic approach than previous studies based on a liquid drop picture. In particular, we have applied the EDF theory, in which nuclear-shell effects are naturally incorporated. We have described both the outer and inner crusts with the same functional, 
thus ensuring a unified and thermodynamically consistent treatment. To assess the impact of the neutron-matter constraint, 
we have employed the set of Brussels-Montreal functionals BSk19, BSk20, and BSk21 \citep{goriely2010}. These functionals were fitted to all 
atomic mass data of nuclei with $Z,N\geq8$ and to different realistic neutron-matter EoSs and spanning different degrees 
of stiffness. For comparison, we have also considered the SLy4 functional \citep{chabanat1998}. All these functionals have been already 
employed to determine the EoS of cold catalysed NSs. 

For the outer crust, we have made use of experimental data from the latest AME 2016~\citep{audi2016} whenever available supplemented with 
theoretical nuclear masses obtained from the HFB method. For the inner crust, we have adopted the 4th order ETF approach with proton shell corrections added perturbatively via the Strutinsky integral theorem. This 
ETFSI method, which was previously applied for catalysed matter~\citep{onsi2008,pearson2012}, is a computationally fast implementation of the HF method. 

We have followed the evolution of an accreted matter element, considering ashes of X-ray bursts made of $^{56}$Fe to compare 
with previous calculations from HZ. As they sink into deeper layers, nuclei undergo a series of electron captures. The first 
two reactions are completely determined by experimental atomic masses, and all models thus predict the same heat sources. 
On the contrary, the location of the sources and the amount heat deposited in the denser region is model dependent. In 
particular, the HFB calculations using SLy4 predict a much larger heat release for the transition from $Z=22$ to $Z=20$
than the other HFB models. This result, however, is the manifestation of an improper description of nuclear shell closure. 
For this reason, SLy4 leads to a significantly larger integrated heat than the other functionals. In contrast to HFB models, 
the heat obtained from the MB model is roughly the same for all reactions due to the lack of nuclear shell effects. 
In any case, the total heat deposited in the outer crust does not exceed $0.2$~MeV. Therefore, the origin of shallow heating 
required to explain the cooling of some SXTs remains unknown, see e.g. ~\cite{parikh2017} and references 
therein. At high enough densities, electron captures are accompanied by neutron emissions thus marking the transition to the inner crust. 
Very accurate analytical formulas for the neutron-drip density and pressure are given in \cite{chamel2016}. Due to the 
constraints imposed on the possible reactions, all models give the same stratification of the outer crust. However, the position 
of the boundary between adjacent layers depends on the model employed. 

In the inner crust, most of the heat is released in electron captures, neutron emissions and absorptions induced by pycnonuclear reactions, 
whereby the proton number of the clusters changes from $Z=16$ to the magic number $Z=14$. Because the shell closure at $Z=14$ 
is a robust prediction of EDF calculations, all functionals give similar amount of the total heat deposited in the inner crust. 
For the same reason, clusters with $Z=14$ are present in large regions of the inner crust. On the contrary, the MB model 
leads to a highly stratified crust. The composition of accreted crust is found to be radically different from that of catalysed crust. In contrast 
to catalysed crust, $Z$ is found to be everywhere smaller than $26$ in accreted crust, and clusters are much lighter. However, the ratios 
$Z/A$ and $A/A_{\rm cell}$ are similar, and less sensitive to fine details of the nuclear structure. 

The total heat deposited in the crust, as predicted by the EDF theory, lies in the range from $1.5$~MeV to $1.7$~MeV. For comparison, 
ETF calculations in the inner crust (with no shell effects) yield much lower values $\sim 0.6$~MeV, thus highlighting 
the importance of nuclear shell effects. The MB model adopted here leads to an intermediate value of $1.2$~MeV due to the inclusion of 
pairing effects. This result is substantially lower than that previously obtained by HZ. The discrepancy arises from the more 
realistic treatment of the neutron emissions and absorptions by clusters in this work. The total crustal heating we have found is 
much lower than that predicted by \cite{steiner2012} within the liquid-drop picture. This stems from an empirical parametrization 
of nuclear shell effects and the adoption by \cite{steiner2012} of the functionals Rs and Gs \citep{fried86} that yield unrealistic neutron-matter EoS. 

The values we find for the crustal heat are somewhat smaller than those  obtained in the recent nuclear network calculations of \cite{LauBeard2018}, and our main heating sources are located deeper. These differences could be attributed to the inclusion by \cite{LauBeard2018} of transitions to excited states and superthreshold electron capture cascades beyond neutron-drip. However, the amounts of cumulated heat predicted by the two approaches appear to be in good agreement at densities $1.6\times 10^{12}$~g~cm$^3$ corresponding to the shallow region of the inner crust. The deviations observed at higher densities where free neutrons are very abundant may thus also come from our consistent description of the neutron-star crust, and in particular the inclusion of the medium modification of nuclear shell effects within the microscopic EDF theory.

\begin{acknowledgements}
We are particularly grateful to J. Margueron for valuable discussions. 
This work was financially supported by F.R.S-FNRS (Belgium), NSERC 
(Canada), and the European Cooperation in Science and Technology (COST) 
actions MP1304 and CA16214. N. C. acknowledges the  support of the Fonds 
de la Recherche Scientifique - FNRS (Belgium) under grant n$^\circ$~CDR-J.0187.16 
and CDR-J.0115.18. 
This work was partially supported by the 
Polish National Science Centre (NCN)  grant no. 2013/11/B/ST9/04528.
\end{acknowledgements}


\begin{appendix}
\section{Tables}
The Tables contain the properties of the reactions
in the crust of accreting NSs for the considered EDFs 
BSk19, BSk20, BSk21, and SLy4, as well as for the MB model. 
We assume that the ashes of 
X-ray bursts consist of pure $^{56}{\rm Fe}$. 
For the EDFs models the atomic number of clusters in the inner crust 
is only approximate (rounded), since in our approach nucleons inside 
clusters and free nucleons are not treated separately. 
For the 
same reason the number $\Delta N$ of emitted neutrons is not specified. 
The pressure and density at which the reactions
take place are denoted by $P$ and $\rho$, respectively.
The relative density jump $\Delta \rho/\rho_j$ and 
the deposited heat (per one accreted nucleon) $Q_j$
associated with the process $j$ are presented in the fifth 
and seventh column respectively. 
$X_n$ is the fraction of free neutrons among  nucleons, 
and $\mu_e$ is the electron Fermi energy,
both in the layer just above the reaction layer.
Pycnonuclear reactios, which lead to the doubled $\acell$, are marked by
horizontal lines.

\begin{table*}[t]
\caption{
Non-equilibrium processes in the crust of an accreting neutron stars assuming
that the ashes of X-ray bursts consist of pure $^{56}{\rm Fe}$ using the EDF BSk21 
(HFB method in the outer crust, ETFSI method in the inner crust). 
}
\label{tab:proc.ETFSI.BSk21}
\begin{center}
\begin{tabular}{lllllrr}
\hline \hline \noalign{\smallskip} $P$ & $\rho$ & Reactions & $X_n$ &
$\Delta \rho/\rho$  &$\mu_e$ &  $q$
\\ (dyn~cm$^{-2}$) & (g~cm$^{-3}$) & & &  & (MeV) 
&(keV)  \\ \hline \noalign{\smallskip}
$6.50\times 10^{26}$ &
$1.38\times 10^9$ & $^{56}\mathrm{Fe}\rightarrow
{}^{56}\mathrm{Cr}-2e^-+2\nu_e $ & 0 & 0.08 & 4.47 & 37.0\\
$1.84\times 10^{28}$ & $1.82\times 10^{10}$
&$^{56}\mathrm{Cr}\rightarrow {}^{56}\mathrm{Ti}-2e^-+2\nu_e$ & 0 & 0.09 & 10.22& 41.2\\
$1.06\times 10^{29}$ & $7.38\times 10^{10}$
 & $^{56}\mathrm{Ti}\rightarrow {}^{56}\mathrm{Ca}-2e^-+2\nu_e$ & 0 & 0.10 & 15.83 & 39.1\\
$3.44\times 10^{29}$ & $1.96\times 10^{11} $
 & $^{56}\mathrm{Ca}\rightarrow
 ^{56}\mathrm{Ar}-2e^-+2\nu_e$ & 0 & 0.11 & 21.22 & 8.1\\
$8.75\times 10^{29}$ & $4.38\times 10^{11} $
 & $^{56}\mathrm{Ar}\rightarrow {}^{55}\mathrm{Cl}+n-e^-+\nu_e$ & 0 & 0.06 & 26.55 & 0\\
$9.40\times 10^{29}$  &   $4.79\times 10^{11}$   &
$^{55}\mathrm{Cl}\rightarrow {}^{53}\mathrm{S}
+\Delta N\cdot n-e^-+2\nu_e$   &   0.05   &   0.06&27.04 &0\\
$1.18\times 10^{30}$  & $6.04\times 10^{11}$   &   ${}^{53}\mathrm{S}\rightarrow
{}^{47}\mathrm{Si}+\Delta N\cdot n-2e^-+2\nu_e$  &  0.15  &0.14  & 28.57 &45.0\\
\hline
\noalign{\smallskip}
$2.54\times 10^{30}$   &   $1.22\times 10^{12}$   &
${}^{48}\mathrm{Si}\rightarrow {}^{30}\mathrm{O}+\Delta N\cdot n-6e^-+2\nu_e
$   &  &  && \\
\noalign{\smallskip}
&&
$ {}^{30}\mathrm{O}+{}^{30}\mathrm{O}\rightarrow {}^{51}\mathrm{Si}+\Delta N\cdot n-2e^-+2\nu_e$&0.54& 0.68 &32.64&908.1\\
\hline
\noalign{\smallskip}
$5.78\times 10^{30}$   &   $3.73\times 10^{12}$   &
${}^{53}\mathrm{Si}\rightarrow {}^{32}\mathrm{O}+\Delta N\cdot n-6e^-+2\nu_e
$   &  &  && \\
\noalign{\smallskip}
&&
${}^{32}\mathrm{O}+{}^{32}\mathrm{O}\rightarrow {}^{62}\mathrm{S}+\Delta N\cdot n$&0.72& 0.23 &35.47&355.9\\
\hline
\noalign{\smallskip}
 $8.69\times 10^{30}$   &   $6.16\times 10^{12}$ &${}^{66}\mathrm{S}\rightarrow {}^{57}\mathrm{Si}+\Delta N\cdot n-2e^-+2\nu_e$ &  
 0.74  &   0.03&37.74&3.5\\
\hline
\noalign{\smallskip}
$3.20\times 10^{31}$   &   $1.65\times 10^{13}$   &
${}^{65}\mathrm{Si}\rightarrow {}^{40}\mathrm{O}+\Delta N\cdot n-6e^-+2\nu_e
$   &  &  && \\
\noalign{\smallskip}
&&
$ {}^{40}\mathrm{O}+{}^{40}\mathrm{O}\rightarrow {}^{76}\mathrm{S}+\Delta N\cdot n$&0.83& 0.05 &43.8&98.2\\
\hline
\noalign{\smallskip}
$1.85\times 10^{32}$  & $7.26\times 10^{13}$   &   ${}^{91}\mathrm{S}\rightarrow
{}^{86}\mathrm{P}+\Delta N\cdot n-e^-+\nu_e$  &  0.81  &0.006  & 69.10 &0\\

\hline
\hline
\end{tabular}
\end{center}
\end{table*}

\begin{table*}[t]
\caption{
Non-equilibrium processes in the crust of an accreting NS assuming
that the ashes of X-ray bursts consist of pure $^{56}{\rm Fe}$ using the EDF BSk20 
(HFB method in the outer crust, ETFSI method in the inner crust). 
}
\label{tab:proc.ETFSI.BSk20}
\begin{center}
\begin{tabular}{lllllrr}
\hline \hline \noalign{\smallskip} $P$ & $\rho$ & Reactions & $X_n$ &
$\Delta \rho/\rho$  &$\mu_e$ &  $q$
\\ (dyn~cm$^{-2}$) & (g~cm$^{-3}$) & & &  & (MeV) 
&(keV)  \\ \hline \noalign{\smallskip}
$6.48\times 10^{26}$ &
$1.38\times 10^9$ & $^{56}\mathrm{Fe}\rightarrow
{}^{56}\mathrm{Cr}-2e^-+2\nu_e $ & 0 & 0.08 & 4.47 & 37.0\\
$1.83\times 10^{28}$ & $1.81\times 10^{10}$
&$^{56}\mathrm{Cr}\rightarrow {}^{56}\mathrm{Ti}-2e^-+2\nu_e$ & 0 & 0.09 & 10.22& 41.2\\
$1.06\times 10^{29}$ & $7.37\times 10^{10}$
 & $^{56}\mathrm{Ti}\rightarrow {}^{56}\mathrm{Ca}-2e^-+2\nu_e$ & 0 & 0.10 & 15.82 & 53.4\\
$3.44\times 10^{29}$ & $1.96\times 10^{11} $
 & $^{56}\mathrm{Ca}\rightarrow
 {}^{56}\mathrm{Ar}-2e^-+2\nu_e$ & 0 & 0.11 & 21.73 & 12.4\\
$9.06\times 10^{29}$ & $4.50\times 10^{11} $
 & $^{56}\mathrm{Ar}\rightarrow  {}^{55}\mathrm{Cl}+ n-e^-+\nu_e$ & 0 & 0.06 & 26.50 & 0\\
$9.32\times 10^{29}$  &   $4.76\times 10^{11}$   &
$^{55}\mathrm{Cl}\rightarrow {}^{54}\mathrm{S}
+\Delta N\cdot n-e^-+2\nu_e$   &   0.04   &   0.06&26.98 &0\\
$1.22\times 10^{30}$  & $6.17\times 10^{11}$   &   $^{54}\mathrm{S}\rightarrow
{}^{48}\mathrm{Si}+\Delta N\cdot n-2e^-+2\nu_e$  &  0.14  &0.13  & 28.78 &49.1\\
\hline
\noalign{\smallskip}
$2.51\times 10^{30}$   &   $1.21\times 10^{12}$   &
$^{48}\mathrm{Si}\rightarrow {}^{30}\mathrm{O}+\Delta N\cdot n-6e^-+2\nu_e
$   &  &  && \\
\noalign{\smallskip}
&&
$^{30}\mathrm{O}+{}^{30}\mathrm{O}\rightarrow {}^{52}\mathrm{Si}-2e^-+2\nu_e$&0.55& 0.71 &32.71&924.8\\
\hline
\noalign{\smallskip}
$5.15\times 10^{30}$   &   $3.49\times 10^{12}$   &
$^{52}\mathrm{Si}\rightarrow {}^{32}\mathrm{O}+\Delta N\cdot n-6e^-+2\nu_e
$   &  &  && \\
\noalign{\smallskip}
&&
$ ^{32}\mathrm{O}+{}^{32}\mathrm{O}\rightarrow {}^{61}\mathrm{S}$&0.73& 0.27&34.98&369.7\\
\hline
\noalign{\smallskip}
 $7.06\times 10^{30}$   &   $5.53\times 10^{12}$ &$^{62}\mathrm{S}\rightarrow {}^{56}\mathrm{Si}+\Delta N\cdot n-2e^-+2\nu_e$ &  
 0.75  & 0.04&36.47&6.3\\
\hline
\noalign{\smallskip}
$2.26\times 10^{31}$   &   $1.32\times 10^{13}$   &
$^{60}\mathrm{Si}\rightarrow {}^{36}\mathrm{O}+\Delta N\cdot n-6e^-+2\nu_e
$   &  &  && \\
\noalign{\smallskip}
&&
$ ^{36}\mathrm{O}+ {}^{36}\mathrm{O}\rightarrow {}^{61}\mathrm{S}$&0.84& 0.07 &40.81&120.7\\
\hline
\noalign{\smallskip}
$1.29\times 10^{32}$  & $5.08\times 10^{13}$   &   ${}^{80}\mathrm{S}\rightarrow
{}^{75}\mathrm{P}+\Delta N\cdot n-e^-+\nu_e$  &  0.83  &0.003  & 61.31 &0\\

\hline
\hline
\end{tabular}
\end{center}
\end{table*}

\begin{table*}[t]
\caption{
Non-equilibrium processes in the crust of an accreting NS assuming
that the ashes of X-ray bursts consist of pure $^{56}{\rm Fe}$ using the EDF BSk19 
(HFB method in the outer crust, ETFSI method in the inner crust). 
}
\label{tab:proc.ETFSI.BSk19}
\begin{center}
\begin{tabular}{lllllrr}
\hline \hline \noalign{\smallskip} $P$ & $\rho$ & Reactions & $X_n$ &
$\Delta \rho/\rho$  &$\mu_e$ &  $q$
\\ (dyn~cm$^{-2}$) & (g~cm$^{-3}$) & & &  & (MeV) 
&(keV)  \\ \hline \noalign{\smallskip}
$6.48\times 10^{26}$ &
$1.38\times 10^9$ & $^{56}\mathrm{Fe}\rightarrow
^{56}\mathrm{Cr}-2e^-+2\nu_e $ & 0 & 0.08 & 4.47 & 37.0\\
$1.83\times 10^{28}$ & $1.81\times 10^{10}$
&$^{56}\mathrm{Cr}\rightarrow {}^{56}\mathrm{Ti}-2e^-+2\nu_e$ & 0 & 0.09 & 10.22& 41.2\\
$1.06\times 10^{29}$ & $7.37\times 10^{10}$
 & $^{56}\mathrm{Ti}\rightarrow {}^{56}\mathrm{Ca}-2e^-+2\nu_e$ & 0 & 0.10 & 15.83 & 62.3\\
$3.44\times 10^{29}$ & $1.96\times 10^{11} $
 & $^{56}\mathrm{Ca}\rightarrow
 ^{56}\mathrm{Ar}-2e^-+2\nu_e$ & 0 & 0.11 & 21.22 & 11.6\\
$9.02\times 10^{29}$ & $4.48\times 10^{11} $
 & $^{56}\mathrm{Ar}\rightarrow {}^{55}\mathrm{Cl}+ n-e^-+\nu_e$ & 0 & 0.06 & 26.74 & 0\\
$9.30\times 10^{29}$  &   $4.75\times 10^{11}$   &
${}^{55}\mathrm{Cl}\rightarrow {}^{54}\mathrm{S}
+\Delta N\cdot n-e^-+2\nu_e$   &   0.04   &   0.06&27.04 &0\\
$1.22\times 10^{30}$  & $6.20\times 10^{11}$   &   ${}^{54}\mathrm{S}\rightarrow
{}^{48}\mathrm{Si}+\Delta N\cdot n-2e^-+2\nu_e$  &  0.14  &0.14  & 28.63 &50.4\\
\hline
\noalign{\smallskip}
$2.48\times 10^{30}$   &   $1.20\times 10^{12}$   &
${}^{48}\mathrm{Si}\rightarrow {}^{30}\mathrm{O}+\Delta N\cdot n-6e^-+2\nu_e
$   &  &  && \\
\noalign{\smallskip}
&&
$ {}^{30}\mathrm{O}+{}^{30}\mathrm{O}\rightarrow {}^{51}\mathrm{Si}+\Delta N\cdot n-2e^-+2\nu_e$&0.54& 0.72 &32.72&932.1\\
\hline
\hline
\noalign{\smallskip}
$4.87\times 10^{30}$   &   $3.38\times 10^{12}$   &
${}^{52}\mathrm{Si}\rightarrow {}^{32}\mathrm{O}+\Delta N\cdot n-6e^-+2\nu_e
$   &  &  && \\
\noalign{\smallskip}
&&
$ {}^{32}\mathrm{O}+{}^{32}\mathrm{O}\rightarrow {}^{61}\mathrm{S}+\Delta N\cdot n$&0.73& 0.28 &34.76&376.8\\
\hline
\noalign{\smallskip}
 $6.36\times 10^{30}$   &   $5.23\times 10^{12}$ &${}^{62}\mathrm{S}\rightarrow {}^{55}\mathrm{Si}+\Delta N\cdot n-2e^-+2\nu_e$ &  
 0.75  &   0.04&35.85&7.4\\
\hline
\noalign{\smallskip}
$1.96\times 10^{31}$   &   $1.21\times 10^{13}$   &
${}^{58}\mathrm{Si}\rightarrow {}^{35}\mathrm{O}+\Delta N\cdot n-6e^-+2\nu_e
$   &  &  && \\
\noalign{\smallskip}
&&
$ {}^{35}\mathrm{O}+{}^{35}\mathrm{O}\rightarrow {}^{68}\mathrm{S}+\Delta N\cdot n$&0.85& 0.07 &39.72&132.0\\
\hline
\noalign{\smallskip}
$8.57\times 10^{31}$  & $3.69\times 10^{13}$   &   $ {}^{75}\mathrm{S}\rightarrow
 {}^{71}\mathrm{P}+\Delta N\cdot n-e^-+\nu_e$  &  0.84  &0.003  & 55.11 &0\\

$1.39\times 10^{32}$  & $5.21\times 10^{13}$   &   $ {}^{72}\mathrm{P}\rightarrow
 {}^{68}\mathrm{Si}+\Delta N\cdot n-e^-+\nu_e$  &  0.85  &0.001  & 60.38 &0\\

\hline
\hline
\end{tabular}
\end{center}
\end{table*}

\begin{table*}[t]
\caption{
Non-equilibrium processes in the inner crust of an accreting NS assuming
that the ashes of X-ray bursts consist of pure $^{56}{\rm Fe}$. ETF model with EDF BSk19.
}
\label{tab:proc.ETF.BSk19}
\begin{center}
\begin{tabular}{lllllrr}
\hline \hline \noalign{\smallskip} $P$ & $\rho$ & Reactions & $X_n$ &
$\Delta \rho/\rho$  &$\mu_e$ &  $q$
\\ (dyn~cm$^{-2}$) & (g~cm$^{-3}$) & & &  & (MeV) 
&(keV)  \\ \hline \noalign{\smallskip}
$9.30\times 10^{29}$  &   $4.75\times 10^{11}$   &
${}^{56}\mathrm{Cl}\rightarrow {}^{54}\mathrm{S}
+\Delta N\cdot n-e^-+\nu_e$   &   0.04   &   0.06&26.97 &0\\
$1.02\times 10^{30}$  & $5.41\times 10^{11}$   &   ${}^{54}\mathrm{S}\rightarrow
{}^{51}\mathrm{P}+\Delta N\cdot n-e^-+\nu_e$  &  0.09  &0.06  & 27.60 &0\\
$1.07\times 10^{30}$  & $5.96\times 10^{11}$   &   ${}^{51}\mathrm{P}\rightarrow
{}^{48}\mathrm{Si}+\Delta N\cdot n-e^-+\nu_e$  &  0.14  &0.07  & 27.90 &0\\
$1.12\times 10^{30}$  & $6.58\times 10^{11}$   &   ${}^{48}\mathrm{Si}\rightarrow
{}^{45}\mathrm{Al}+\Delta N\cdot n-e^-+\nu_e$  &  0.20  &0.07  & 28.18 &0\\
$1.16\times 10^{30}$  & $7.27\times 10^{11}$   &   ${}^{45}\mathrm{Al}\rightarrow
{}^{42}\mathrm{Mg}+\Delta N\cdot n-e^-+\nu_e$  &  0.25  &0.07  & 28.41 &0\\
$1.22\times 10^{30}$  & $8.08\times 10^{11}$   &   ${}^{42}\mathrm{Mg}\rightarrow
{}^{39}\mathrm{Na}+\Delta N\cdot n-e^-+\nu_e$  &  0.31  &0.08  & 28.66 &0\\
$1.28\times 10^{30}$  & $9.05\times 10^{11}$   &   ${}^{39}\mathrm{Na}\rightarrow
{}^{36}\mathrm{Ne}+\Delta N\cdot n-e^-+\nu_e$  &  0.37  &0.08  & 28.91 &0\\
$1.36\times 10^{30}$  & $1.02\times 10^{12}$   &   ${}^{36}\mathrm{Ne}\rightarrow
{}^{32}\mathrm{F}+\Delta N\cdot n-e^-+\nu_e$  &  0.42  &0.09  & 29.19 &0\\
\hline
\noalign{\smallskip}
$1.44\times 10^{30}$   &   $1.16\times 10^{12}$   &
${}^{32}\mathrm{F}\rightarrow {}^{29}\mathrm{O}+\Delta N\cdot n-e^-+\nu_e
$   &  &  && \\
\noalign{\smallskip}
&&
$ {}^{29}\mathrm{O}+{}^{29}\mathrm{O}\rightarrow {}^{57}\mathrm{S}+\Delta N\cdot n$&0.49& 0.09 &29.40&384.1\\
\hline\noalign{\smallskip}

$1.77\times 10^{30}$  & $1.47\times 10^{12}$   &   ${}^{57}\mathrm{S}\rightarrow
{}^{54}\mathrm{P}+\Delta N\cdot n-e^-+\nu_e$  &  0.52  &0.05  & 30.57 &0\\
$1.89\times 10^{30}$  & $1.62\times 10^{12}$   &   ${}^{54}\mathrm{P}\rightarrow
{}^{50}\mathrm{Si}+\Delta N\cdot n-e^-+\nu_e$  &  0.55  &0.05  & 30.86 &0\\
$2.04\times 10^{30}$  & $1.79\times 10^{12}$   &   ${}^{51}\mathrm{Si}\rightarrow
{}^{47}\mathrm{Al}+\Delta N\cdot n-e^-+\nu_e$  &  0.58  &0.05  & 31.19 &0\\
$2.25\times 10^{30}$  & $2.01\times 10^{12}$   &   ${}^{47}\mathrm{Al}\rightarrow
{}^{44}\mathrm{Mg}+\Delta N\cdot n-e^-+\nu_e$  &  0.61  &0.05  & 31.63 &0\\
$2.50\times 10^{30}$  & $2.27\times 10^{12}$   &   ${}^{44}\mathrm{Mg}\rightarrow
{}^{41}\mathrm{Na}+\Delta N\cdot n-e^-+\nu_e$  &  0.63  &0.05  & 32.09 &0\\
$2.82\times 10^{30}$  & $2.60\times 10^{12}$   &   ${}^{41}\mathrm{Na}\rightarrow
{}^{39}\mathrm{Ne}+\Delta N\cdot n-e^-+\nu_e$  &  0.66  &0.05  & 32.59 &0\\
$3.25\times 10^{30}$  & $3.00\times 10^{12}$   &   ${}^{39}\mathrm{Ne}\rightarrow
{}^{34}\mathrm{F}+\Delta N\cdot n-e^-+\nu_e$  &  0.69  &0.04  & 33.13 &0\\
\hline\noalign{\smallskip}

$3.84\times 10^{30}$   & $3.53\times 10^{12}$   &
${}^{34}\mathrm{F}\rightarrow {}^{31}\mathrm{O}+\Delta N\cdot n-e^-+\nu_e
$   &  &  && \\
\noalign{\smallskip}
&&
$ {}^{31}\mathrm{O}+{}^{31}\mathrm{O}\rightarrow {}^{60}\mathrm{S}+\Delta N\cdot n$&0.73& 0.04 &33.78&183.8\\
\hline\noalign{\smallskip}

$5.38\times 10^{30}$  & $4.65\times 10^{12}$   &   ${}^{61}\mathrm{S}\rightarrow
{}^{58}\mathrm{P}+\Delta N\cdot n-e^-+\nu_e$  &  0.74  &0.02  & 35.58 &0\\
$6.24\times 10^{30}$  & $5.26\times 10^{12}$   &   ${}^{58}\mathrm{P}\rightarrow
{}^{55}\mathrm{Si}+\Delta N\cdot n-e^-+\nu_e$  &  0.76  &0.02  & 36.29 &0\\
$7.38\times 10^{30}$  & $6.03\times 10^{12}$   &   ${}^{55}\mathrm{Si}\rightarrow
{}^{52}\mathrm{Al}+\Delta N\cdot n-e^-+\nu_e$  &  0.77  &0.02  & 37.12 &0\\
$9.05\times 10^{30}$  & $7.09\times 10^{12}$   &   ${}^{52}\mathrm{Al}\rightarrow
{}^{48}\mathrm{Mg}+\Delta N\cdot n-e^-+\nu_e$  &  0.78  &0.02  & 38.22 &0\\
$1.13\times 10^{31}$  & $8.45\times 10^{12}$   &   ${}^{49}\mathrm{Mg}\rightarrow
{}^{45}\mathrm{Na}+\Delta N\cdot n-e^-+\nu_e$  &  0.80  &0.01  & 39.44 &0\\
$1.51\times 10^{31}$  & $1.05\times 10^{13}$   &  ${}^{46}\mathrm{Na}\rightarrow
{}^{42}\mathrm{Ne}+\Delta N\cdot n-e^-+\nu_e$  &  0.81  &0.01  & 41.20 &0\\
$2.12\times 10^{31}$  & $1.35\times 10^{13}$   &   ${}^{43}\mathrm{Ne}\rightarrow
{}^{40}\mathrm{F}+\Delta N\cdot n-e^-+\nu_e$  &  0.82  &0.01  & 43.37 &0\\
\hline\noalign{\smallskip}

$3.29\times 10^{31}$   & $1.87\times 10^{13}$   &
${}^{41}\mathrm{F}\rightarrow {}^{37}\mathrm{O}+\Delta N\cdot n-e^-+\nu_e
$   &  &  && \\
\noalign{\smallskip}
&&
$ {}^{37}\mathrm{O}+{}^{37}\mathrm{O}\rightarrow {}^{70}\mathrm{S}+\Delta N\cdot n$&0.84& 0.001 &46.64&73.9\\
\hline\noalign{\smallskip}


$7.48\times 10^{31}$  & $3.34\times 10^{13}$   &   ${}^{74}\mathrm{S}\rightarrow
{}^{70}\mathrm{P}+\Delta N\cdot n-e^-+\nu_e$  &  0.84  &0.003  & 54.46 &0\\

\hline
\hline
\end{tabular}
\end{center}
\end{table*}

\begin{table*}[t]
\caption{
Non-equilibrium processes in the crust of an accreting NS assuming that the ashes 
of X-ray bursts consist of pure $^{56}{\rm Fe}$ using the EDF SLy4.
(HFB method in the outer crust, ETFSI method in the inner crust). 
}
\label{tab:proc.ETFSI.SLy}
\begin{center}
\begin{tabular}{lllllrr}
\hline \hline \noalign{\smallskip} $P$ & $\rho$ & Reactions & $X_n$ &
$\Delta \rho/\rho$  &$\mu_e$ &  $q$
\\ (dyn~cm$^{-2}$) & (g~cm$^{-3}$) & & &  & (MeV) 
&(keV)  \\ \hline \noalign{\smallskip}
$6.48\times 10^{26}$ &
$1.38\times 10^9$ & $^{56}\mathrm{Fe}\rightarrow
^{56}\mathrm{Cr}-2e^-+2\nu_e $ & 0 & 0.08 & 4.47 & 37.0\\
$1.83\times 10^{28}$ & $1.81\times 10^{10}$
&$^{56}\mathrm{Cr}\rightarrow {}^{56}\mathrm{Ti}-2e^-+2\nu_e$ & 0 & 0.09 & 10.22& 41.2\\
$1.06\times 10^{29}$ & $7.37\times 10^{10}$
 & $^{56}\mathrm{Ti}\rightarrow {}^{56}\mathrm{Ca}-2e^-+2\nu_e$ & 0 & 0.10 & 15.83 & 111.5\\
$4.50\times 10^{29}$ & $2.40\times 10^{11} $
 & $^{56}\mathrm{Ca}\rightarrow
 {}^{56}\mathrm{Ar}-2e^-+2\nu_e$ & 0 & 0.11 & 22.70 & 13.2\\
$9.29\times 10^{29}$ & $4.58\times 10^{11} $
 & $^{56}\mathrm{Ar}\rightarrow {}^{55}\mathrm{S}+ n-2e^-+2\nu_e$ & 0 & 0.12 & 26.95 & 12.0\\
$1.18\times 10^{30}$  & $6.03\times 10^{11}$   &   ${}^{53}\mathrm{S}\rightarrow
{}^{47}\mathrm{Si}+\dz\cdot n-2e^-+2\nu_e$  &  0.14  &0.06  & 28.61 &60.9\\
\hline
$2.26\times 10^{30}$   &   $1.11\times 10^{12}$   &
${}^{47}\mathrm{Si}\rightarrow {}^{29}\mathrm{O}+\dz\cdot n-6e^-+2\nu_e
$   &  &  && \\
\noalign{\smallskip}
&&
$ {}^{29}\mathrm{O}+{}^{29}\mathrm{O}\rightarrow {}^{50}\mathrm{Si}+\dz\cdot n-2e^-+2\nu_e$&0.16& 0.69 &33.57&890.0\\
\hline
\hline
$4.82\times 10^{30}$   &   $3.27\times 10^{12}$   &
${}^{51}\mathrm{Si}\rightarrow {}^{31}\mathrm{O}+\dz\cdot n-6e^-+2\nu_e
$   &  &  && \\
\noalign{\smallskip}
&&
$ {}^{31}\mathrm{O}+{}^{31}\mathrm{O}\rightarrow {}^{60}\mathrm{S}+\dz\cdot n$&0.54& 0.26 &38.15&353.8\\
\hline
 $7.42\times 10^{30}$   &   $5.55\times 10^{12}$ &${}^{61}\mathrm{S}\rightarrow {}^{55}\mathrm{Si}+\dz\cdot n-2e^-+2\nu_e$
 &  0.73  &   0.03&37.75&8.3\\
\hline
$2.26\times 10^{31}$   &   $1.28\times 10^{13}$   &
${}^{59}\mathrm{Si}\rightarrow {}^{36}\mathrm{O}+\dz\cdot n-6e^-+2\nu_e
$   &  &  && \\
\noalign{\smallskip}
&&
$ {}^{36}\mathrm{O}+{}^{36}\mathrm{O}\rightarrow {}^{69}\mathrm{S}+\dz\cdot n$&0.74& 0.06 &47.66&114.0\\
\hline
\noalign{\smallskip}
$1.23\times 10^{32}$  & $4.78\times 10^{13}$   &   $ {}^{78}\mathrm{S}\rightarrow
 {}^{74}\mathrm{P}+\dz\cdot n-e^-+\nu_e$  &  0.83  &0.003  & 61.31 &0\\

$1.94\times 10^{32}$  & $5.21\times 10^{13}$   &   ${}^{74}\mathrm{P}\rightarrow
{}^{69}\mathrm{Si}+\dz\cdot n-e^-+\nu_e$  &  0.84  &0.001  & 67.08 &0\\

\hline
\hline
\end{tabular}
\end{center}
\end{table*}

\begin{table*}[t]
\caption{
Non-equilibrium processes in the crust of an accreting NS assuming
that the ashes of X-ray bursts consist of pure $^{56}{\rm Fe}$. Mackie-Baym model adapted
to the scheme presented in this paper.
}
\label{tab:proc.MB}
\begin{center}
\begin{tabular}{lllllrr}
\hline \hline \noalign{\smallskip} $P$ & $\rho$ & Reactions & $X_n$ &
$\Delta \rho/\rho$  &$\mu_e$ &  $q$
\\ (dyn~cm$^{-2}$) & (g~cm$^{-3}$) & & &  & (MeV) 
&(keV)  \\ \hline \noalign{\smallskip}
$6.51\times 10^{26}$ &
$1.38\times 10^9$ & $^{56}\mathrm{Fe}\rightarrow
^{56}\mathrm{Cr}-2e^-+2\nu_e $ & 0 & 0.08 & 4.47 & 36.8\\
$9.57\times 10^{27}$ & $1.11\times 10^{10}$
&$^{56}\mathrm{Cr}\rightarrow {}^{56}\mathrm{Ti}-2e^-+2\nu_e$ & 0 & 0.09 & 8.69& 35.8\\
$1.15\times 10^{29}$ & $7.85\times 10^{10}$
 & $^{56}\mathrm{Ti}\rightarrow {}^{56}\mathrm{Ca}-2e^-+2\nu_e$ & 0 & 0.10 & 16.15 & 47.2\\
$4.75\times 10^{29}$ & $2.50\times 10^{11} $
 & $^{56}\mathrm{Ca}\rightarrow
 ^{56}\mathrm{Ar}-2e^-+2\nu_e$ & 0 & 0.11 & 22.99 & 46.1\\
$1.07\times 10^{30}$ & $5.09\times 10^{11} $
 & $^{56}\mathrm{Ar}\rightarrow {}^{55}\mathrm{S}+ n-2e^-+2\nu_e$ & 0 & 0.12 & 27.63 & 19.0\\
$1.46\times 10^{30}$  &   $7.22\times 10^{11}$   &
$^{52}\mathrm{S}\rightarrow {}^{46}\mathrm{Si}
+6n-2e^-+2\nu_e$   &   0.17   &   0.13&30.28 &52.7\\
$1.58\times 10^{30}$  & $8.66\times 10^{12}$   &   $^{46}\mathrm{Si}\rightarrow
^{40}\mathrm{Mg}+6n-2e^-+2\nu_e$  &  0.29  &0.15  & 30.68 &55.7\\
$1.74\times 10^{30}$   &   $1.07\times 10^{12}$   &
$^{40}\mathrm{Mg}\rightarrow {}^{34}\mathrm{Ne}+6n-2e^-+2\nu_e
$ & 0.39 & 0.16 & 31.11 & 59.2 \\
\hline
$1.97\times 10^{30}$   &   $1.36\times 10^{12}$   &
$^{34}\mathrm{Ne}\rightarrow {}^{28}\mathrm{O}+6n-2e^-+2\nu_e
$   &  &  && \\
\noalign{\smallskip}
&&
$ ^{28}\mathrm{O}+ {}^{28}\mathrm{O}\rightarrow {}^{54}\mathrm{S}$&0.52& 0.18 &31.48&380.7\\
\hline
 $2.38\times 10^{30}$   &   $1.86\times 10^{12}$ &$^{54}\mathrm{S}\rightarrow {}^{48}\mathrm{Si}+6n-2e^-+2\nu_e$ &  0.57  &   0.08&32.47&24.3\\
$2.75\times 10^{30}$   &  $2.24\times 10^{12}$   &   $^{48}\mathrm{Si}\rightarrow
^{42}\mathrm{Mg}+6n-2e^-+2\nu_e$    &0.63   &0.09  &  32.92& 25.0\\
$3.31\times 10^{30}$   &   $2.81\times 10^{12}$   &  $^{42}\mathrm{Mg}\rightarrow
^{36}\mathrm{Ne}+6n-2e^-+2\nu_e$  &0.68 & 0.10 & 33.41&25.8\\
\hline
\noalign{\smallskip}
$4.23\times 10^{30}$   &   $3.67\times 10^{12}$   &
$^{36}\mathrm{Ne}\rightarrow {}^{30}\mathrm{O}+6n-2e^-+2\nu_e
$   &  &  && \\
\noalign{\smallskip}
&&
$ ^{30}\mathrm{O}+ {}^{30}\mathrm{O}\rightarrow {}^{58}\mathrm{S}$&0.74& 0.09 &33.89&193.8\\
\hline
\noalign{\smallskip}
$6.09\times 10^{30}$  & $5.23\times 10^{12}$  & $^{58}\mathrm{S}\rightarrow
^{52}\mathrm{Si}+6n-2e^-+2\nu_e$  &  0.77  &0.04  & 35.84 &10.2\\
$7.87\times 10^{30}$  &  $6.51\times 10^{12}$  &  $^{52}\mathrm{Si}\rightarrow
 ^{46}\mathrm{Mg}+6n-2e^-+2\nu_e$&
0.79  &  0.04  &36.67 &10.4\\
$1.08\times 10^{31}$   &  $8.48\times 10^{12}$   &   $^{46}\mathrm{Mg}\rightarrow
^{40}\mathrm{Ne}+6n-2e^-+2\nu_e$    &0.82   &0.04  &  37.64& 10.4\\
\hline
\noalign{\smallskip}
$1.62\times 10^{31}$   &   $1.18\times 10^{13}$   &
$^{42}\mathrm{Ne}\rightarrow {}^{34}\mathrm{O}+6n-2e^-+2\nu_e
$   &  &  && \\
\noalign{\smallskip}
&&
$ ^{34}\mathrm{O}+ {}^{34}\mathrm{O}\rightarrow {}^{50}\mathrm{S}$&0.85& 0.03 &38.82&99.4\\
\hline
\noalign{\smallskip}

\hline
\noalign{\smallskip}
$3.32\times 10^{31}$  & $2.00\times 10^{13}$  & $^{70}\mathrm{S}\rightarrow
^{62}\mathrm{Si}+8n-2e^-+2\nu_e$  &  0.86  &0.01  & 44.06 &3.9\\
$5.49\times 10^{31}$  &  $2.88\times 10^{13}$  &  $^{68}\mathrm{Si}\rightarrow
 ^{60}\mathrm{Mg}+8n-2e^-+2\nu_e$&
0.87 &  0.009  &47.21 &3.6\\
\noalign{\smallskip}

\hline
\hline
\end{tabular}
\end{center}
\end{table*}

\end{appendix}

\end{document}